\documentclass[preprint,amsmath,amssymb,superscriptaddress,longbibliography,authoryear,5p,twocolumn,times]{elsarticle}
\usepackage[titletoc,toc,title]{appendix}
\journal{NeuroImage}
\usepackage{hyperref}
\hypersetup{colorlinks = true, linkcolor = blue, anchorcolor = blue, citecolor = blue, pdfstartpage = 1, filecolor = red, urlcolor = black}
\usepackage{graphicx}
\usepackage{amsmath}
\usepackage{amssymb}
\usepackage{amsfonts}
\usepackage{mathrsfs}
\usepackage{subfigure}
\usepackage{float} 
\usepackage{soul}
\usepackage{marginnote}
\usepackage{fancyhdr}
\renewcommand{\headrulewidth}{0pt}
\usepackage{xcolor}
\usepackage{xspace}
\usepackage[english]{babel}
\def\cite{\citep}
\newcommand{\mpar}[1]{}
\newcommand{\mpars}[1]{}
\newcommand{\mpart}[1]{}

\newcommand{\mnote}[1]{}
\newcommand{\mnotes}[1]{}

\newcommand{\new}[1]{{\color{black} #1}}
\newcommand{\news}[1]{{\color{black} #1}}
\newcommand{\newt}[1]{{\color{black} #1}}
\newcommand{\units}[1]{\,\mathrm{#1}}
\def\const{\xspace\mathrm{const}\xspace}
\def\x{{\bf x}}
\def\k{{\bf k}}
\def\q{{\bf q}}

\def\D{{\cal D}}

\def\O{{\cal O}}
\def\lp{\left(}
\def\rp{\right)}
\def\lb{\left[}
\def\rb{\right]}

\def\Dinf{D_{\infty}}

\def\w{\omega}
\def\d{\mbox{d}}

\def\be{\begin{equation}}
\def\ee{\end{equation}}
\def\bea{\begin{eqnarray}}
\def\eea{\end{eqnarray}}
\def\tauex{{\tau_{\rm ex}}}

\renewcommand{\eqref}[1]{\textup{{Eq.~(\ref{#1})}}}
\newcommand{\figref}[1]{\textup{{Fig.~\ref{#1}}}}

\begin{document}

%\preprint{APS/123-QED}

\title{In vivo observation and biophysical interpretation of time-dependent diffusion in human cortical gray matter}
%\title{Time-dependent diffusion coefficient and kurtosis in human gray matter}% Force line breaks with \\

\author[cbi,cai2r]{Hong-Hsi Lee\fnref{fn1}\corref{cor}}
\author[cbi,cai2r]{Antonios Papaioannou\fnref{fn1}}
\author[cbi,cai2r]{Dmitry S. Novikov}
\author[cbi,cai2r]{Els Fieremans}
% \address{Center of Biomedical Imaging, Department of Radiology, New York University School of Medicine,  New York, NY 10016, USA}
\fntext[fn1]{These authors contributed equally to the work.}
\cortext[cor]{Corresponding author: Honghsi.Lee@nyulangone.org}
\address[cbi]{Center for Biomedical Imaging, Department of Radiology, New York University School of Medicine, New York, NY, USA}
\address[cai2r]{Center for Advanced Imaging Innovation and Research (CAI2R), New York University School of Medicine, New York, NY, USA}
\date{\today}% It is always \today, today,

\begin{abstract}
\new{The dependence of the diffusion MRI signal on the diffusion time $t$ is a hallmark of tissue microstructure at the scale of the diffusion length. 
%, and has been observed in human gray matter using diffusion MRI. 
%Here, we provide a more comprehensive study by measuring the
Here we measure the time-dependence of the mean diffusivity $D(t)$ and mean kurtosis $K(t)$ in cortical gray matter and in 25 gray matter sub-regions, in 10 healthy subjects. Significant diffusivity and kurtosis time-dependence is observed for $t=21.2$-100 ms, and is characterized by a power-law tail $\sim t^{-\vartheta}$ with dynamical exponent $\vartheta$.
To interpret our measurements, we systematize the relevant scenarios and mechanisms for diffusion time-dependence in the brain. Using effective medium theory formalisms, we derive an exact relation between the power-law tails in $D(t)$ and $K(t)$.  
The estimated power-law dynamical exponent $\vartheta\simeq1/2$ in both $D(t)$ and $K(t)$ is consistent with one-dimensional diffusion in the presence of randomly positioned restrictions along neurites. 
We analyze the short-range disordered statistics of synapses on axon collaterals in the cortex, and perform one-dimensional Monte Carlo simulations of diffusion restricted by permeable barriers with a similar randomness in their placement, to confirm the $\vartheta=1/2$ exponent.
In contrast, the K\"arger model of exchange is less consistent with the data since it does not capture the diffusivity time-dependence, and the estimated exchange time from $K(t)$ falls below our measured $t$-range.  
%and its fit to kurtosis time-dependence yields an exchange time of $11\,$ms, much shorter than previously found values, and below the shortest $t$ in our measurements. 
Although we  cannot exclude exchange as a contributing factor, we argue that structural disorder along neurites is mainly responsible for the observed time-dependence of diffusivity and kurtosis. 
Our observation and theoretical interpretation of the $t^{-1/2}$ tail in $D(t)$ and $K(t)$ alltogether establish the sensitivity of a macroscopic  MRI signal to micrometer-scale structural heterogeneities along neurites in human gray matter in vivo.}

\end{abstract}
\maketitle
%\keywords{}
%\abbreviations{}

\section{Introduction}
The effect of varying the diffusion time $t$ on the diffusion MRI (dMRI) signal has been 
%The  time-dependence of the diffusion coefficient in the neuronal tissue, 
studied in neuronal tissue since the 1990's \cite{Horsfield_time_d,beaulieu1996,stanisz1997,assaf2000,Does2003}, and has been increasingly used for quantifying neuronal microstructure \cite{nilsson2009,Kunz2013,Pyatigorskaya,novikov2014revealing,Wu2014,Burcaw2015,Fieremans2016,Palombo2016pnas,jespersen2018,Lee2018}, as reviewed by \citet{review-nbm}. Such investigations are of interest, as they are complementary to traditional $q$-space imaging at fixed $t$, widely used in clinical studies. Furthermore, measurement of the time-dependent dMRI signal offers a direct probe to restrictions at the scale of the diffusion length $L(t) = \sqrt{\langle x^2(t)\rangle} \sim1-10\units{\mu m}$ (defined as root-mean-squared molecular displacement), and in principle allows one to classify \cite{novikov2014revealing} and quantify the corresponding microstructural features in brain non-invasively \cite{Latour15021994,Barazany2009,Burcaw2015,Fieremans2016,DeSantis2016,Benjamini2016,Lee2018}. 

% Should we include a short paragraph, or just one sentence on what has been done so far in white matter? mention axon diameter mapping and axial time-dependence  due to axon caliber variation?
So far, the time-dependence of the dMRI signal has been most often studied by measuring the diffusion coefficient $D(t)=\langle x^2\rangle/2t$. 
In gray matter, frequency dependence of the diffusion coefficient, $D(\omega)$, was previously reported in rat cortical areas using oscillating gradient spin echo techniques (OGSE) between 20--1000 Hz \cite{Does2003}, \mpar{R3.2}\new{in the mouse brain between 50--150 Hz \citep{aggarwal2012ogse}}, \newt{and in the human brain between 25-50 Hz \citep{baron2014ogse}}. Similar OGSE techniques revealed $D(\omega)$ in adult mouse cerebellum \citep{Wu2014} \newt{and in human brain white matter \citep{arbabi2020ogse}}. In addition, the diffusion kurtosis, $K(t)=\langle x^4\rangle/\langle x^2\rangle^2-3$ \cite{kiselev2010cumulant,JensenKurt}, was shown to have a non-monotonic behavior at short times in rat cortex, between 2 and 29 ms, using both conventional PGSE (pulsed gradient spin echo) and OGSE techniques \cite{Pyatigorskaya}. The same PGSE and OGSE techniques were used to study $D(t)$ and $K(t)$ in healthy and injured mouse brains \cite{Jiangyang2018}\news{, and ex vivo cuprizone-treated mouse brains \citep{aggarwal2020kurtosis}}. 
Using numerical simulations, the finer microstructure of dendrites has been studied by constructing artificial spines along dendrites and investigating the time-dependence of an intra-dendritic diffusion coefficient \cite{PALOMBO2017}. However, the time-dependence of the dMRI signal in cortical areas of the human brain in vivo has not yet been investigated. 

Here we \new{measure $D(t)$ and $K(t)$ in vivo in human cortical gray matter for $t=21.2-100 \units{ms}$ using a standard clinical PGSE sequence at fixed echo time on} a clinical scanner. To interpret our measurements, we consider the effect of coarse graining of the structural disorder by diffusion, and the effect of water exchange, on $D(t)$ and $K(t)$. 

\new{{\it Structural disorder} causes the time-dependence of diffusion metrics \cite{novikov2014revealing}.} With increasing diffusion time $t$, water molecules coarse-grain the underlying micro-architecture over increasing length scales $L(t)$, such that, qualitatively, a medium (e.g., a tissue compartment) can be effectively viewed  as a set of domains of the size $\sim L(t)$, each with a different local diffusion coefficient \new{$D(\x_0)|_{L(t)}$} \cite{review-nbm}. While averaging over $L\to\infty$ would completely homogenize the medium, resulting in asymptotically Gaussian diffusion with effective $D|_{t\to\infty}\equiv \Dinf$, at {\it finite} $t$ and $L(t)$, such coarse-graining is incomplete. This gradual approach to Gaussian diffusion manifests itself in a characteristic inverse power-law time-dependence of $D(t)$ within a given tissue compartment \cite{novikov2014revealing,Burcaw2015,Fieremans2016,jespersen2018}.  
Likewise, the higher-order cumulants, such as $K(t)$, acquire time-dependence \cite{novikov2010emt,Burcaw2015,Dhital2017} due to incomplete coarse-graining (as a measure of the residual inhomogeneity of the effective medium). 
The same underlying physics of coarse-graining results in the power-law  behavior $t^{-\vartheta}$ of both $D(t)$ and $K(t)$ at long diffusion times with identical power-law exponents \cite{Burcaw2015,Dhital2017}. 
 
 A competing mechanism for time-dependent kurtosis, $K(t)$, may be the {\it exchange} between compartments --- relevant even when diffusion in each of them can be already considered Gaussian at a given diffusion time \cite{KM}. The way to think about the diffusion physics in this situation is to imagine that  coarse-graining has already completed in each %non-exchanging%
 compartment, with slow exchange remaining between compartments. In this case, the overall $D(t)$ remains time-independent (as a weighted average of Gaussian compartment diffusivities), while the kurtosis decreases to zero asymptotically as $1/t$ as a manifestation of exchange. 
 
As these two mechanisms result in distinct time-dependencies, studying both $D(t)$ and $K(t)$ with a focus on their asymptotic behavior at long $t$ offers ways to probe the relevant microstructural degrees of freedom --- e.g., the presence of intra-compartmental non-Gaussianity connected to incomplete coarse-graining, and the related disorder universality class and/or effective dimensionality, as well as the importance of exchange between compartments, and the relative role of intra- and inter-compartmental kurtosis. 

%However, a non zero diffusion kurtosis may rise from multiple anisotropic, Guassian compartments due to exchange or alternatively, from the presence of microstructure making diffusion non-Gaussian and giving rise to time-dependence. Time-dependent diffusion experiments could therefore disentangle this ambiguity.

%Performing such studies and connecting the dynamics of diffusion with the structure may reveal some of the salient features of microstructure in the cortex, such as spine density in dendrites, which may be connected with several neuropsychiatric disorders \cite{penzes2011dendritic}.

%Here we report empirically the $D(t)$ and $K(t)$ in cortical gray matter areas of the human brain at diffusion times between $t=21.2-100 \units{ms}$ using a standard clinical PGSE sequence at fixed echo time. 
%Our experimental findings reveal a distinct time-dependence in kurtosis, whereas the diffusion coefficient remains approximately flat at the time-scale of the experiment. 
\new{The outline of this work is as follows. In Section \ref{theory}, we put the relevant mechanisms, such as the disorder coarse-graining picture, and the exchange picture, into  overarching context of a model-selection tree for the brain microstructure, Fig.~\ref{Fig:Hierarchy}. We then present our experimental setup (Section \ref{sec:methods}) and results (Section \ref{sec:results}). 
To interpret our experimental findings, we explore relevant branches of the model-selection tree, by deriving exact relations, Eqs.~(\ref{xi}) and (\ref{K-xi}), between power-law tails in $D(t)$ and $K(t)$ (Section \ref{theory} and \ref{sec:app-power}), and by performing Monte Carlo simulations of one-dimensional diffusion in the presence of short-range disorder, with restrictions mimicking those along synapses on axon collaterals in the cortex based on our analysis of microscopy data (Section \ref{sec:results} and \ref{sec:app-perm}). 
We discuss our theoretical and experimental results in Section \ref{discussion}.}  
% what esle did we do? structural analysis? simulations?
%The observed time-dependence may be interpreted with several models, i.e. diffusion in one-dimension in the intra-neurite space, diffusion in two-dimensions from the intra- and extra-neurite space, or exchange between the two compartments. 
%In what follows, we investigate the validity of those models against our experimental data. In addition, 
%In particular, we connect the structure of axons in cortical gray matter with the observed $K(t)$, by undertaking a detailed analysis of their structural characteristics. %In what follows, we introduce various models of diffusion starting from a simple case of Gaussian diffusion in a single compartment. More complex cases are discussed such as exchange between Gaussian compartments and ultimately introducing the case of non-Gaussian diffusion in the presence of microstructure. These models are introduced in the form of a selection tree based on empirical observation of time-dependence in the diffusion coefficient and kurtosis.

\section{Theory}
\label{theory}

%%%%%%%%%%%%%%%%%%%%%%%%%%%%%%%%%%%%%
\begin{figure*}[t]
\centering
	\includegraphics[width=1.0\textwidth]{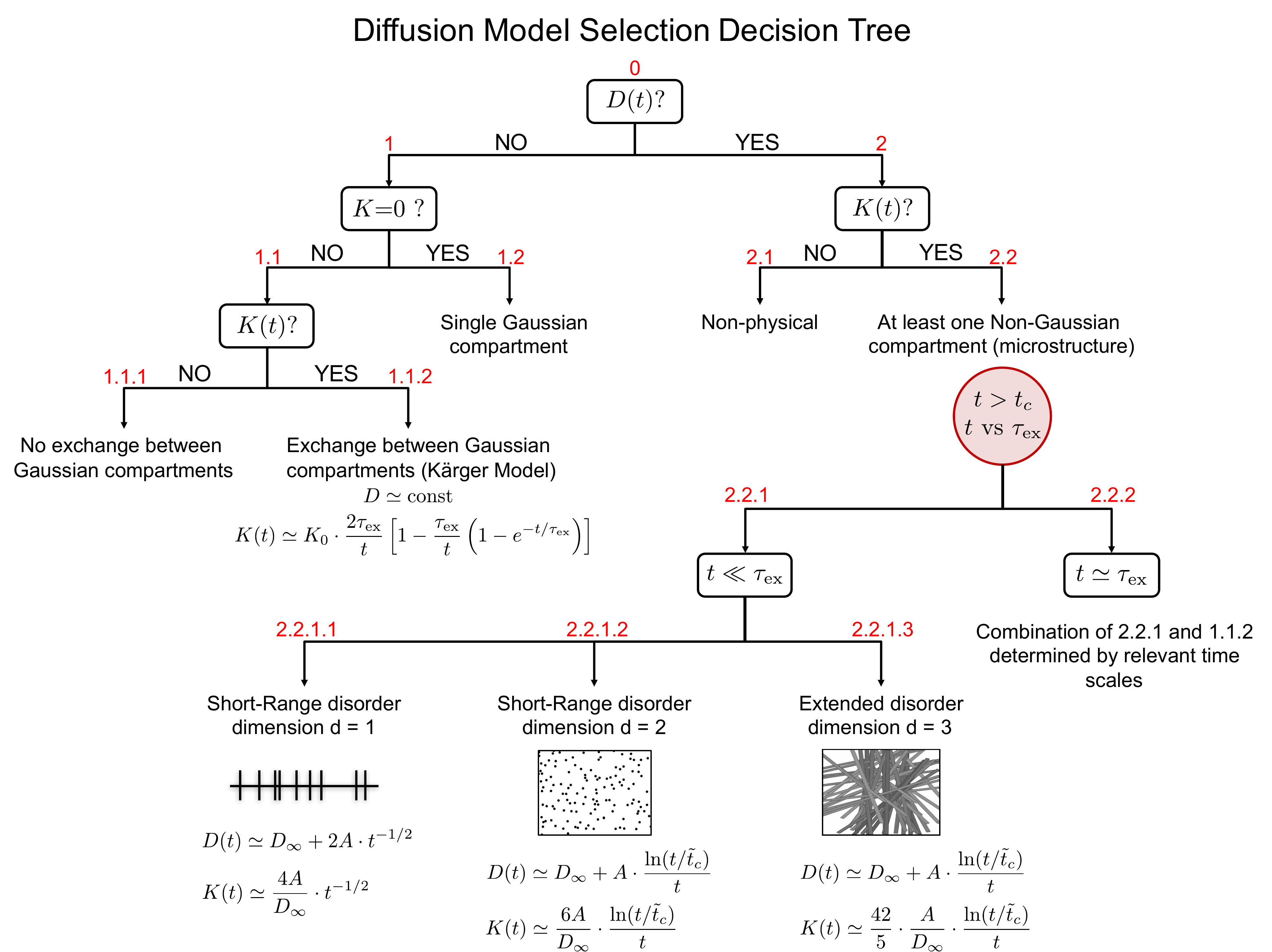}
	\caption[]{\textbf{Model-selection tree for the brain microstructure.}\mnote{R4.8} The main criterion for moving down the tree is the time-dependence of the diffusion coefficient and kurtosis. The picture of non-exchanging Gaussian compartments, such as the Standard Model  of impermeable stick-like axons embedded in a Gaussian extra-axonal space, falls within Node 1.1.1, whereas the K{\"a}rger model  of exchange between Gaussian compartments in Node 1.1.2, cf. Eq.~(\ref{KurtosisKM}). 
If a time-dependent diffusion coefficient is observed, the long-time scaling of $D(t)$ and $K(t)$ can be used to determine the structural disorder universality class, \new{some of which are sketched within Node 2.2.1 and in \figref{Fig:cartoon}.} 
The  effects of exchange add to the time-dependence of $K(t)$ and compete with the disorder coarse-graining effects (Node 2.2.2).
}
	\label{Fig:Hierarchy}
\end{figure*}
%%%%%%%%%%%%%%%%%%%%%%%%%%%%%%%%%%%%

In this Section we introduce the hierarchy of models that describe the connection between $D(t)$ and $K(t)$ with the various compartments and microstructure types that are likely to be present in the brain (Fig. \ref{Fig:Hierarchy}). The resulting ``selection tree" summarizes the various diffusion models from top to bottom, such as Gaussian diffusion in exchanging compartments, or diffusion in the presence of microstructure. Moving down the tree's nodes is decided based on the presence or absence of time-dependence of $D$ and/or $K$. The selection tree illustrates that  measuring the fourth-order cumulant (kurtosis) is essential to reveal the physical picture of the system of interest. Note that subsections in this Section are numbered based on the selection tree nodes in Fig.~\ref{Fig:Hierarchy}. 

\subsection*{Node 1: Gaussian compartments}

This is a broad class of phenomena where coarse-graining over the microstructure in each compartment has already happened, so that for all practical purposes, all compartments can be considered as homogeneous at the scale of $L(t)$ probed by the measurement. In this case, neither compartment diffusivity depends on time, and therefore, the overall $D=\const$.

\subsection*{Node 1.2: Single Gaussian compartment}
The simplest case  is molecular diffusion in a uniform medium, i.e., a Gaussian compartment. This results in no time-dependence in the diffusion coefficient and  zero kurtosis (as well as all higher-order cumulants); examples are pure liquids. 
%In this case $D(t)\simeq D_0$ and $K\simeq0.0$ $\forall$ $t$; where $D_0$ is the water diffusivity at the corresponding temperature.  
% why not going with the commented out line?

\subsection*{Node 1.1.1: Non-exchanging Gaussian compartments}

A non-zero kurtosis indicates the presence of multiple compartments (which can be anisotropic). 
This physical picture underpins, e.g., the so-called Standard Model of diffusion in white matter \cite{review-nbm}, 
generalizing a number of previous works  \cite{kroenke2004nature,jespersen2007modeling,Jespersen2010,FIEREMANS2011177,noddi,sotiropoulos2012,jensen2016,baydiff,rotinv}, some of which have been also applied to gray matter. In this case, one compartment consists of so-called ``sticks" \cite{kroenke2004nature,jespersen2007modeling}, i.e., narrow impermeable cylinders of finite diffusivity in the direction of the principal axis and negligible transverse diffusivity --- modeling neurites. Other compartments then include the extra-neurite space as a locally Gaussian compartment, and possibly CSF as yet another distinct Gaussian compartment. 
In all such model variations, the diffusion coefficient and kurtosis (tensors) are time-independent. In the general case of $n$ non-interacting Gaussian compartments with fractions $p_i$ and (directional) diffusivities $D_i$, the overall diffusivity 
\begin{equation}
 \overline{D} = \sum\limits_{i=1}^n p_iD_i \,, \quad \sum_{i=1}^n p_i = 1\,, 
 \label{D_non_interacting}
 \end{equation} 
 and kurtosis 
 \begin{equation} \label{K0}
  K_0 = 3\frac{\mathrm{var}\{D\}}{\overline{D}^2} \,, \quad 
  \mathrm{var}\{D\}=\sum\limits_{i=1}^np_i(D_i-\overline{D})^2
 \end{equation}
were given by \citet{DKI}. 

\subsection*{Node 1.1.2: Exchanging Gaussian compartments}
\label{SR_KM}

The presence of time-dependence in $K(t)$ with no time-dependence in $D$ 
indicates exchange between Gaussian compartments, 
while the residual, non-Gaussian {\it intra}-compartmental effects are negligible. 
In this ``adiabatic exchange" regime, the  
K{\"a}rger model (KM) \cite{KARGER19881} originally developed for chemical solutions has been shown to apply to complex tissue environments \cite{KM}.
In this case, the diffusivity is time-independent and given by Eq.~(\ref{D_non_interacting}), whereas kurtosis decays on  the exchange time scale
$t\sim \tauex$  \cite{DKI,KM}:
\begin{equation}
K(t)=K_0\cdot \frac{2\tau_{\rm{ex}}}{t}\lb 1-\frac{\tau_{\rm{ex}}}{t}(1-e^{-t/\tau_{\rm{ex}}})\rb,
\label{KurtosisKM}
\end{equation}
where $K_0\equiv K(t)|_{t=0}$ is  given by Eq.~(\ref{K0}) above, exemplifying that exchange effects can be neglected for $t\ll \tauex$. 
Conversely, for $t\gg\tauex$, kurtosis approaches its limit $K(t)|_{t\to\infty} = 0$ of a Gaussian medium asymptotically as $\sim 1/t$. 
\new{Finite-pulse PGSE generalization of  \eqref{KurtosisKM} was  found in the $t\ll\tau_\text{ex}$ limit \citep{ning2018karger}.} \mpar{R2.6}
  
The presence of non-exchanging Gaussian compartments within a voxel would add a constant $K(t)|_{t\to\infty} = K_{\infty}$ to Eq.~(\ref{KurtosisKM}), whereas the $t$-dependent part (\ref{KurtosisKM}) would then describe exchanging compartments (with a suitably redefined $K_0$). 
This candidate behavior will be compared to our experimental findings in Section~\ref{modelFit} and Fig.~\ref{Fig:Mod_Sel} below. 

\mpart{R2.1}\newt{The KM is exact for only few scenarios, such as for diffusion \textit{along} highly aligned packed cylinders that are exchanging in between. However, in most geometries of biological tissues, diffusion in each compartment is non-Gaussian, violating the assumptions of the KM. Therefore, this model usually serves as an asymptotic approximation at long times, when the micro-geometry of each compartment is almost coarse-grained as an effective medium of Gaussian diffusion. For example, KM could describe the asymptotic behavior of the diffusion transverse to packed cylinders at long times, especially for slow exchange regimes \citep{KM}. }

%we investigate this model of exchange against our experimental data. 
%In addition, we investigate a modified K{\"a}rger Model, $K_{\rm{mod}}(t) = K(t)+ K_{\infty}$. The added constant drives Eq.~(\ref{KurtosisKM}) to $K(t)|_{t\to\infty} = K_{\infty}$ rather than to $K(t)|_{t\to\infty} = 0$, pointing towards addition compartments exchanging with the system of interest at the time scales of the experiment.

% \begin{equation*}
%     K(t)\simeq K_0\left(1-\frac{t'}{3\tau_\text{ex}}\right)\,,\quad
%     t' = \frac{t^3-t^2\delta+\tfrac{2}{3}\delta^2t-\tfrac{4}{21}\delta^3}{(t-\delta/3)^2}\,,
% \end{equation*}
% where $t'\simeq t-\delta/3$ for $t\gg \delta$, 
 %the full functional form for all $t$ has not been explored yet.}

\subsection*{Node 2: Intra-compartmental microstructure effects}
Node 2 of Fig.~\ref{Fig:Hierarchy} corresponds to the time-dependence of the diffusion coefficient, or $D(t)$. 
In the absence of flow, $D(t)$, to the best of our knowledge, can only originate from the presence of  microstructure, cf. Sections 1.9 and 2 of the review article by \citet{review-nbm} for a detailed discussion. 
Incomplete coarse-graining of the microstructure manifests itself in non-Gaussian diffusion; this results in time-dependence of {\it both} $D(t)$ and of non-zero higher-order cumulants, i.e., $K(t)$ and beyond.

\subsection*{Node 2.1: Non-physical case}

We are unaware of a physical system where the diffusivity is time-dependent and the kurtosis does not depend on time (at any time scale): 
Physically, the former would indicate that the coarse-graining is not over, while the latter corresponds to  complete coarse-graining. This contradiction suggests  checking the processing pipeline with respect to parameter estimation biases but also the calibration of the MRI pulse sequence that is being used. For those types of contradicting results, pulse sequence calibration using an ice\new{-water} phantom \cite{Malyarenko2016,FIEREMANS201839} is recommended.

\subsection*{Node 2.2: Diffusion in the presence of microstructure}

For this general case of both $D(t)$ and $K(t)$ being time-dependent, we will assume the range of diffusion times 
$t > t_c$ to exceed the correlation time $t_c$ corresponding to diffusing past the correlation length $l_c = L(t_c)$ of tissue microstructure in a given compartment. This assumption is reasonable for the brain, since the size of typical structural ``features"  within the neuropil (spines, boutons, axon and dendrite diameters) is about 1\,$\mu$m, corresponding to $t_c\sim 1\,$ms, while our diffusion time range is at least an order of magnitude greater. 

While the neuropil generally dominates the cellular volume \cite{chklovskii-optimization2002}, 
we note that this assumption may be invalid in certain individual cortical layers with notable density of neuronal soma.
% , and assume here that our voxels are too large to have such soma-rich domains dominating the physics. Otherwise,
\mpar{R4.2}
\new{In this case, the diffusion inside the neuronal bodies should also be modeled, generally leading to the soma contributions to $D(t)$ and $K(t)$ both decreasing as $1/t$ for $t\gg t_D = R^2/D|_{t\to0}$ as originating from a closed compartment of \new{soma radius} $\sim R$. For $t \lesssim t_D$, $K(t)$ from soma would {\it increase} with $t$; as below we observe that $K(t)$ decreases monotonically, and such short-$t$ contribution seems undetectable in the overall $K(t)$ of our in vivo measurements. } 

%%%%%%%%%%%%%%%%%%%%%%%%%%%%%%%%%%%%%
\begin{figure*}[th!!]
\centering
	\includegraphics[width=0.7\textwidth]{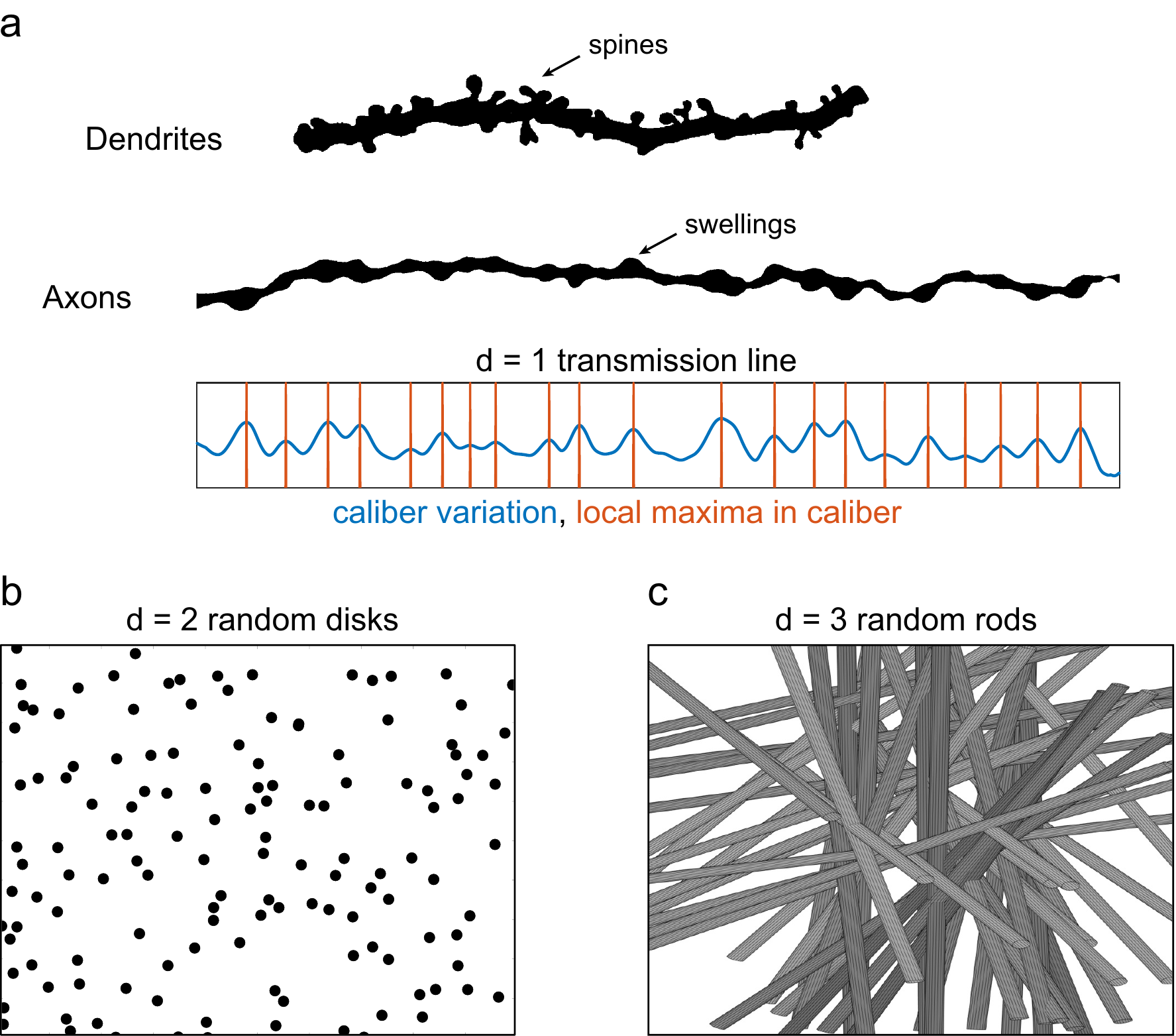}
	\caption[]{\textbf{Cartoon representation of mapping the complex microstructure onto simpler systems.}\mnote{R4.1} a) Mapping of dendrites \news{\citep{woolley1990dendrite}} and axons \new{\citep{Shepherd6340}} into a $d=1$ dimensional transmission line with barriers of permeability $\kappa$ \new{(Node 2.2.1.1 in \figref{Fig:Hierarchy}). Here shows an example of caliber variation (blue) along an axon, and the local maxima (red) in caliber are identified as microstructural inhomogeneity along the axon.} b) A system of randomly distributed disks in $d=2$ dimensions \new{(Node 2.2.1.2 in \figref{Fig:Hierarchy})}. c) A $d=3$ dimensional system of random rods (Node 2.2.1.3 in \figref{Fig:Hierarchy}). 
	\news{The panel a) is adapted with permission from \citep{woolley1990dendrite}, Copyright 1990 Society for Neuroscience, and \citep{Shepherd6340}, Copyright 2002 National Academy of Sciences.}}
	\label{Fig:cartoon}
\end{figure*}
%%%%%%%%%%%%%%%%%%%%%%%%%%%%%%%%%%%%%

\subsection*{Node 2.2.1: Effects of intra-compartmental microstructure; no exchange between compartments, 
$t_c \ll t\ll\tauex$}

Coarse-graining the microstructure in a given compartment past the correlation length, i.e., $L(t)\gg l_c$, results in distinct power-law tails  \cite{novikov2014revealing,review-nbm} in the {\it instantaneous} diffusion coefficient for this compartment, 
\begin{equation} 
D_{\rm inst}(t) \equiv \frac12 \partial_t \langle x^2(t) \rangle \simeq D_{\infty} + A \cdot t^{-\vartheta} \,.
\label{eq:instDiffusion}
\end{equation} 
Here, the dynamical exponent 
\be \label{vartheta}
\vartheta = (p+d)/2
\ee
is related to the compartment's spatial dimensionality $d$, and to the disorder universality class, 
defined in terms of the structural exponent $p$ describing long-range density fluctuations \new{$n(\x_0)$} of the microstructure via its
power spectrum:  \mpar{R4.1}\new{
\begin{equation}
\label{eq:Correlator}
\Gamma(k)\equiv \int_V\! \d\x\, e^{-i\k\cdot\x} \langle n(\x_0+\x)n(\x_0) \rangle_{\x_0}=\frac{\lvert \tilde n(\k)\rvert^2}{V} \sim k^p \,, \quad k\to0 \,.
\end{equation}}
Here $V$ is the compartment volume (or length in $d=1$)\new{, and $\tilde n(\k)$ is the Fourier transform of $n(\x_0)$}. 
In other words, coarse-graining the structurally disordered microstructure $n(\x_0)$ over the diffusion length $L(t)$ corresponds to probing the variance $\Gamma(k)$ of the structural fluctuations $n(\x_0)$ at the corresponding wave vector $k\sim 1/L(t)$. In this way, measuring the diffusive dynamics enables probing the degree of spatial correlations of microstructural building blocks. 
The coefficient $A$ in Eq.~(\ref{eq:instDiffusion}) is proportional to that in front of $k^p$ in Eq.~(\ref{eq:Correlator}); we can therefore say that $\Gamma(k) \propto A\cdot k^p$ as $k\to 0$.

The typically measured {\it cumulative} diffusion coefficient 
\be \label{D=Dinst}
D(t) = \frac1{2t} \langle x^2(t)\rangle = \frac1t \int_0^t\!  D_{\rm inst}(t')\d t' 
\ee
will have the same power-law scaling 
\be
D(t)  \simeq D_{\infty} + c_D\cdot t^{-\vartheta}\,, \quad c_D = {A \over 1-\vartheta} \,, \quad \mbox{for} \quad \vartheta < 1 \,,
\label{eq:cuDiffusion}
\ee
and will approach $D_\infty$ as $\sim 1/t$ for $\vartheta > 1$ \cite{novikov2014revealing}.
The borderline case of $\vartheta=1$ yields the $\ln(t/\tilde{t}_c)/t$ behavior \cite{Burcaw2015}
\begin{equation}
D(t)\simeq D_{\infty} + A\frac{\ln(t/\tilde t_c)}{t}\,, \quad \tilde t_c \sim \mbox{max} \{t_c, \ \delta\} \,, \quad \vartheta=1 \,,
\label{eq:instDiffusionSR2}
\end{equation} 
where $\delta$ is PGSE pulse width.
\new{The $1/(1-\vartheta)$ divergence in $c_D$ as $\vartheta\to 1$ can be attributed to $\ln\tilde t_c$, as described in \ref{sec:app-power}.
The $\ln (t/\tilde{t}_c)/t$ behavior is applicable when $t\gg\tilde{t}_c$. For wide gradient pulses, i.e., $t\gtrsim\delta\gg t_c$, this functional form is generalized to \citep{Burcaw2015} 
\begin{equation*}
    \frac{\ln(t/\tilde{t}_c)}{t}\to \frac{\ln(t/\delta)+\tfrac{3}{2}}{t-\delta/3}\,.
\end{equation*}
We will use this generalized form below for our finite-$\delta$ measurements.}\mpar{R2.6}

\new{The central theoretical result of this work is the general relation between the power law tails in $D(t)$ and $K(t)$ for any $p$ and $d$.} Specifically, the same power-law exponent $\vartheta$ appears in the kurtosis for $t\gg t_c$: 
\begin{equation}
K(t) \simeq  c_K\cdot t^{-\vartheta}, \quad \vartheta < 1 \,, 
\label{eq:cuKurtosis}
\end{equation} 
% where $K_\infty\equiv 0$ 
for a single compartment ( $K(t)\to0$ in $t\to\infty$ limit as diffusion asymptotically becomes Gaussian). 
% \news{In GM, non-zero $K_\infty$ could be caused by an extra dot compartment with negligible diffusivity and non-vanishing signals at even high b-values; the Rician noise floor, if not fully corrected, could have the same effect on the $K_\infty$.}
\mpar{R4.1}\new{Moreover, the dimensionless ratio $\xi$ of the tails $K(t)$ and $\lb D(t)-D_\infty\rb/D_\infty$, is {\it exactly} given in terms of 
$p$ and $d$  (\ref{sec:app-power}): 
\begin{equation} \label{xi}
 \xi(p,d) \equiv \frac{c_K}{c_D/D_\infty} =  6 \cdot \left[ \left( 2 + \frac{p(3p+d-4)}{2(d+2)} \right) \cdot \frac{1}{2-\vartheta} - 1 \right].
\end{equation}
The borderline case $\vartheta=1$ has the same $\ln(t/\tilde{t}_c) /t$ behavior as in \eqref{eq:instDiffusionSR2}, with $c_K$ formally diverging as $1/(1-\vartheta)$, \ref{sec:app-power}; their ratio $\xi|_{p+d=2}$ remains regular, defining the amplitude of the $\ln(t/\tilde{t}_c)/t$ tail in kurtosis: 
\be \label{K-xi}
K(t) \simeq \xi|_{p+d=2} \cdot \frac{A}{D_\infty} \frac{\ln(t/\tilde t_c)}{t}\,,  \quad \vartheta = 1\,.
\ee}

% If a number of compartments are present, their power-law tails will compete, such that the one with the smallest $\vartheta$ will dominate in the overall diffusivity $D(t)$ and overall kurtosis $K(t)$ (the slowest to decay at long $t$), and the asymptotic kurtosis value $K_\infty$ will be given in terms of the variance of the long-time limits  in the non-exchanging compartment diffusivities, cf. \new{Node 1.1.1} and Eq.~(\ref{K0}) with $D_i \to D_{\infty,i}$. 

The coefficients $A$, $c_D$ and $c_K$, and the constant $D_{\infty}$ of Eqs.~(\ref{eq:cuDiffusion}) and (\ref{eq:cuKurtosis}) are 
non-universal, i.e., they depend on the microstructural features such as compartment volume fractions, membrane permeability, characteristic sizes of microstructural building blocks, and their exact placements, see Eqs.~(\ref{Dinf-SR})--(\ref{eq:A}) below for an example.  
Conversely, the power-law exponent (\ref{vartheta}) \new{and the tail ratio (\ref{xi})} are {\it universal}, i.e., they take distinct values for a given compartment depending on its structural universality class and dimensionality, and are thereby robust to continuous changes of  tissue parameters and biological variability. 

\new{
Beside the theoretical generality of the results (\ref{xi}) and (\ref{K-xi}), we note that practically, within the limited range of diffusion times in actual experiments, any above functional forms of $D(t)$ and $K(t)$ can fit the measured time-dependence well. The exact result for the tail ratio allows us to further narrow down the choice between the models of structural disorder, instead of just relying on the goodness-of-fit for $\vartheta$. 
Furthermore, as we see,  the same tail in $D(t)$ can originate from distinct $p$ and $d$, in which case the knowledge of an exact tail ratio is essential. 
}
\mpar{R1.1}

In Sections \ref{Gamma_axonsGM}, \ref{modelFit} and \ref{structure} below, we will analyze the structural correlations and the temporal scaling laws (\ref{eq:cuDiffusion}) and (\ref{eq:cuKurtosis}) for the microstructure  in gray matter.  
%To aid our analysis, it is useful to have in mind a few likely microstructural arrangements which lead to the candidate behaviors (\ref{eq:cuDiffusion}) and (\ref{eq:cuKurtosis}). 
Below we consider relevant microstructural arrangements: 

\new{
\begin{itemize}
\item Node 2.2.1.1, diffusion inside narrow long neurites (axons and dendrites), restricted by spines, beads, shafts and other heterogeneities with local density $n(x_0)$, Fig.~\ref{Fig:cartoon}a. \mpar{R4.1}
Coarse-graining over the diffusion length $L(t)$ exceeding both the typical distance between the restrictions and the neurite diameter (so that the diffusion can be considered one-dimensional) maps the diffusion in a 3-dimensional neurite onto a one-dimensional diffusion with a diffusivity $D(x_0)$ smoothly varying on the scale $\gtrsim L(t)$, whose long-range fluctuations  mimic those of $n(x_0)$.  
In Section~\ref{structure} we will show that the power spectrum $\Gamma(k)$ of $n(x_0)$, Eq.~(\ref{eq:Correlator}), 
is characterized by the structural exponent $p=0$ as $k\to 0$. In dimension $d=1$, this yields $\vartheta = 1/2$ \cite{novikov2014revealing} and the ratio $\xi(0,1)=2$ \cite{Dhital2017}, such that
\be \label{c'-theta}
D(t) - D_\infty \simeq 2A \cdot t^{-1/2} \,, \quad K(t) \simeq  {4A\over D_\infty} \cdot t^{-1/2} \,.
%c_K|_{p=0,\ d=1} = {2c_D \over D_\infty} = {4A\over D_\infty} \,.
\ee

\item Node 2.2.1.2, diffusion in the extra-neurite space transverse to a coherent randomly-packed fiber bundle, \figref{Fig:cartoon}b.
\citet{Burcaw2015} showed that such a neuronal tract is characterized by short-range disorder, exponent $p=0$ in dimension $d=2$, yielding $\vartheta = 1$, $D(t)$ described by Eq.~(\ref{eq:instDiffusionSR2}), and the ratio $\xi(0,2)=6$, such that 
\begin{equation}
K(t) \simeq \frac{6A}{D_{\infty}} \frac{\ln(t/\tilde t_c)}{t} \,, \quad p=0 \ \ \mbox{in} \ \ d=2\,.
\label{eq:instKurtosisSR2}
\end{equation} 

\item \new{Node 2.2.1.3}, 
diffusion in the extra-neurite space of randomly placed and oriented neurites embedded in a $d=3$-dimensional space, Fig.~\ref{Fig:cartoon}c. 
This is an example of {\it extended disorder} (``random rods") \cite{novikov2014revealing}, for which exponent $p=-1$, such that  structural fluctuations diverge. While $D(t)$ has the same form (\ref{eq:instDiffusionSR2}) as in Node 2.2.1.2, Eq.~(\ref{K-xi}) yields  different $\xi(-1,3) = 42/5$, i.e., 
\begin{equation}
K(t) \simeq \frac{42}5\, \frac{A}{D_{\infty}}\frac{\ln(t/\tilde t_c)}{t} \,, \quad p=-1 \ \ \mbox{in} \ \ d=3 \,.
\label{eq:instKurtosisExtend}
\end{equation} 
\end{itemize}
The last two nodes exemplify the fact that both disorder classes -- short-range in $d=2$ and extended in $d=3$ -- create qualitatively similar restrictions to diffusion, governed by the dynamical exponent $\vartheta = 1$.
They can be further distinguished by the tail ratio of $K(t)$ and $D(t)$, Eqs.~(\ref{xi})--(\ref{K-xi}). 
%The kurtosis has a similar tail to that in $D(t)$,  with the coefficient $c_K$ in front of $\ln (t/\tilde{t}_c)/t$ being numerically different in the two cases. 
}

If a number of compartments are present, their power-law tails will compete, such that the one with the smallest $\vartheta$ will dominate in the overall diffusivity $D(t)$ and overall kurtosis $K(t)$ (the slowest to decay at long $t$), and the asymptotic kurtosis value $K_\infty$ will be given in terms of the variance of the long-time limits  in the non-exchanging compartment diffusivities, cf. Node 1.1.1 and Eq.~(\ref{K0}) with $D_i \to D_{\infty,i}$. \news{In this case, the power-law scaling in \eqref{eq:cuKurtosis} is extended as
\be \label{eq:cuKurtosis-inf}
K(t) \simeq K_\infty + c_K\cdot t^{-\vartheta}\,,\quad \vartheta<1\,,
\ee
and the generalization of \eqref{xi} is further discussed in \ref{sec:app-power}.}

\subsection*{Node 2.2.2: Competition between intra-compartmental microstructure and inter-compartmental exchange, 
 $t_c \ll \tauex \lesssim t$}

An interesting case emerges when, while coarse-graining occurs in each compartment, molecules can hop between the compartments: that is, the exchange begins to interfere with nontrivial intra-compartmental diffusion. 
While this case has not been studied quantitatively,  qualitative considerations were given in Appendix F of \cite{Burcaw2015}, arguing  that the adiabatic exchange does not alter the dynamical exponent. 
In this picture, the overall diffusivity $D(t)$ will scale with the slowest compartmental dynamical exponent 
$\vartheta$ provided that $\vartheta < 1$, 
and such intra-compartmental $t^{-\vartheta}$ scaling will also dominate in the overall $K(t)$, since its asymptotic decrease  due to the exchange happens with a  power-law tail $K(t) \sim 1/t$, cf. Eq.~(\ref{KurtosisKM}), that decays faster than that in Eq.~(\ref{eq:cuKurtosis}). 
The logarithmic singularity for $\vartheta=1$ (if such exponent is dominant) will also hold in both $D(t)$ and $K(t)$, cf. Eq.~(\ref{K-xi}). 
Finally, for $\vartheta>1$, similar considerations predict that 
$D(t)$ and $K(t)$ will decrease as $1/t$ with non-universal coefficients, which will not be immediately related to each other (contrary to Eq.~(\ref{xi})), since the one in $D(t)$ would be dominated by the non-universal short-time behavior of $D_{\rm inst}(t)$ according to Eq.~(\ref{D=Dinst}) \cite{PhysRevE.96.061101}, 
while that in $K(t)$ will have the admixture of exchange, cf. Eq.~(\ref{KurtosisKM}).

For white matter, the intra-extra axonal exchange rate $\tauex^{-1}$ was found to range  between $0.3 - 1.8\,$s$^{-1}$  \cite{Lampinen2017}. For neurons and glial cells grown on polysterene beads, the exchange time was  recently estimated to be 
$\tauex\approx 115\,$ms  \cite{Donghan2014}.
\new{In live and fixed excised neonatal mouse spinal cord, \citet{williamson2019spine} observed the water exchange rate $\sim$ 100 s$^{-1}$ between membrane structures and free environments. }
Measurement for diffusion times of the order of or exceeding 100\,ms may thereby be affected by the physics of exchange.  

\begin{figure*}[t]
\centering
	\includegraphics[width=1.00\textwidth]{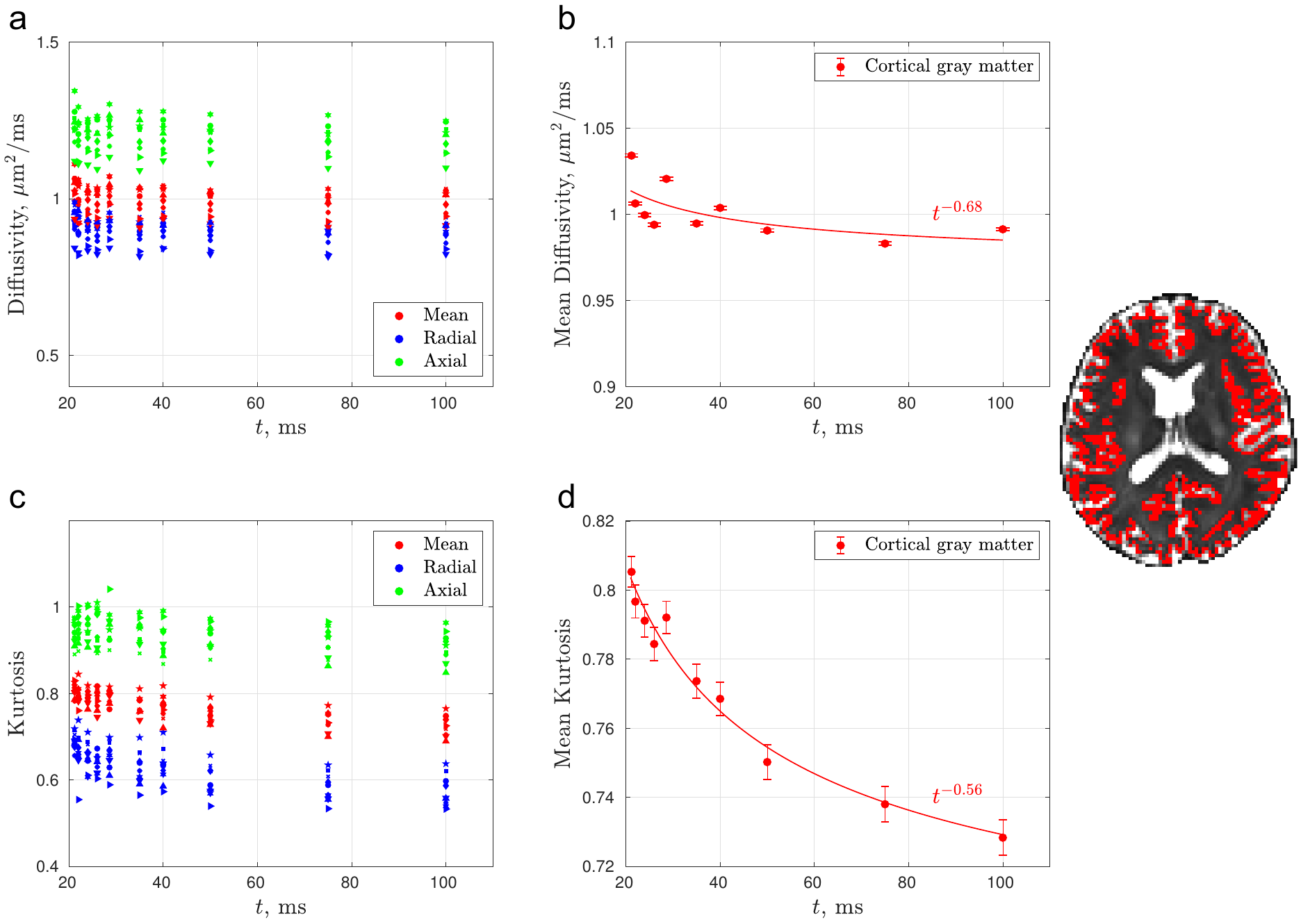}
	\caption[]{\textbf{Time-dependence of diffusion metrics in human cortical gray matter.} \mnote{R1.2\\R1.4\\R4.3}\new{Diffusivity reveals a weak and noisy time-dependence, whereas diffusional kurtosis reveals a strong and distinct time-dependence (Both are significant with P-values $<$ 0.05).} a) Time-dependence of axial, radial and mean diffusivity for all 10 subjects. b) time-dependence of mean diffusivity averaged among all subjects. c) time-dependence of axial, radial and mean kurtosis for all subjects. d) time-dependence of mean kurtosis averaged over all subjects. \news{Solid lines in b) and d) are fits based on the power-law scaling (\ref{eq:cuDiffusion}) and (\ref{eq:cuKurtosis-inf}) in Node 2.2.1.} Right panel: Cortical gray matter ROI shown on a $b=0$ image.}
	\label{Fig:Main_Fig}
\end{figure*}
%%%%%%%%%%%%%%%%%%%%%%%%%%%%%%%%%%%%%

%%%%%%%%%%%%%%%%%%%%%%%%%%%%%%%%%%%%%
\begin{figure*}[t]
\centering
	\includegraphics[width=1.00\textwidth]{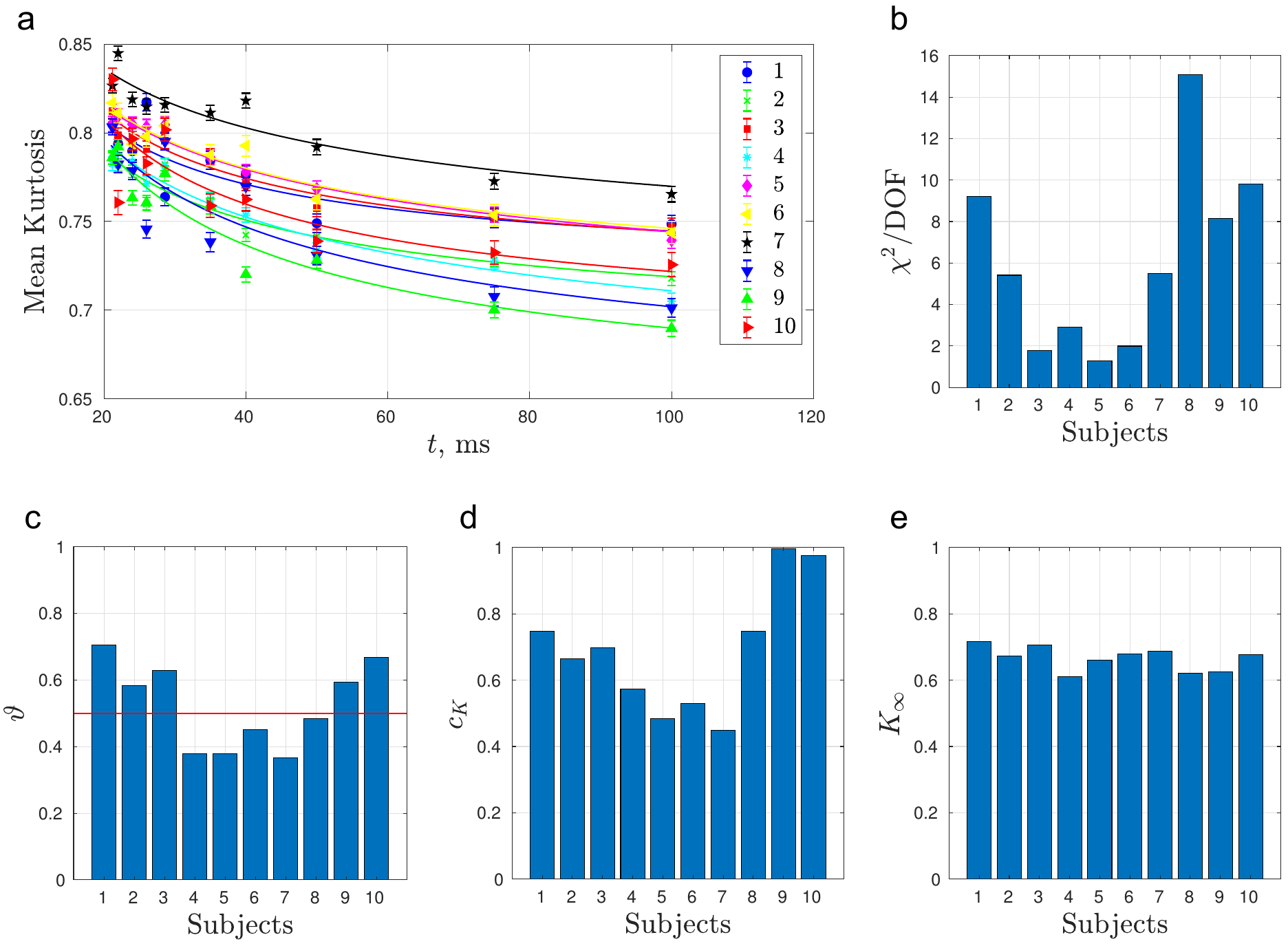}
	\caption[]{\textbf{The time-dependence of diffusion kurtosis \news{in cortical gray matter} appeared consistent between subjects.} \mnote{R1.5\\R4.6}a) Time-dependence of mean kurtosis for all subjects. b) Goodness of fit according to Eq.~(\ref{eq:cuKurtosis-inf}) for all subjects scanned in this work. 
% 	Note that subject \#1 had a $\chi^2/DOF\simeq 12$. 
	It is observed that averaging over all subjects improves the quality of the fit substantially. c) Fitted dynamical exponent $\vartheta$ for all subjects scanned in this work. d) Fitted coefficient $c_K$ was found to have \new{moderate} variation between subjects. e) Fitted $K_{\infty}$ for all subjects scanned in this work. }
	\label{Fig:Scaling_subjects_statistics}
\end{figure*}
%%%%%%%%%%%%%%%%%%%%%%%%%%%%%%%%%%%%%

%%%%%%%%%%%%%%%%%%%%%%%%%%%%%%%%%%%%%%%%%%%%%%%%%
\section{Methods}
\label{sec:methods}

\subsection{Acquisition}
Diffusion MRI was performed on 10 healthy volunteers (7 males and 3 females) ranging between 23 to 30 years old on a Siemens Prisma (3T) system after obtaining a consent which was approved by the Institutional Review Board. A monopolar \new{PGSE} (Siemens WIP 511E) diffusion weighting sequence was used for acquiring \new{diffusion-weighted images (DWIs) of} four b-shells ($b=[0.1, 0.4, 1.0, 1.5]$ ms/$\mu$m$^2$) along 64 directions in total. In addition, 2 $b=0$ images were acquired, one with phase encoding according to anterior-posterior (AP), the same as the DWIs, and one addition according to posterior-anterior (PA) for distortion correction. % check the number of b = 0 images
The diffusion time, identified as $t=\Delta$ in the PGSE sequence, was varied as $21.2-22-24-26-28.6-35-40-50-75-100$ ms, all with the same gradient pulse duration $\delta=15$ ms. (The approximate equivalence of $t$ with $\Delta$, with its precision determined by $\delta$, is explained in Section~2.3 of the review by \citet{review-nbm}.) 
The remaining experimental parameters of the sequence are detailed below: $\rm{TE}=150$ ms, $\rm{TR}=5000$ ms,  resolution = 2.0$\times$2.0$\times$2.0 mm$^3$. A slab of \new{15} slices was acquired and was aligned parallel to the anterior commissure (AC) - posterior commissure (PC) line. The total scan time for each subject was approximately one hour.  

The sequence was calibrated using an ice-water phantom \cite{Malyarenko2016} at $0^{\circ}$C, resulting in $D_0=1.1$ $\rm{\mu m^2/ms}$ and $K=0.01$ over a diffusion time range $t=21.2-100$ ms \news{(\figref{Fig:Ice_waterphantom} in Supplementary Material)}, verifying that there is no artificial time-dependence induced in the diffusion coefficient or kurtosis by the pulse sequence (Supplemental Fig.~S1). An MPRAGE image was also acquired with resolution = 1.0$\times$1.0$\times$1.0 mm$^3$, TE $=2.7$ ms, TR $=2100$ ms, and used for \new{the gray matter }segmentation. %more parameters need to be provided for the MPR

\subsection{Data processing} % can you add citation to Ben's DESIGNER paper?

The processing pipeline of the diffusion weighted images \cite{ADESARON2018532} included noise reduction using MPPCA (mrtrix dwidenoise) \cite{VERAART2016394} resulting a signal-to-noise ratio (SNR) $\approx35$ in $b=0$ images, Gibbs ringing removal (mrtrix mrdegibbs) \cite{MRM:MRM26054}, \new{correction of susceptibility-induced distortion (FSL topup) \citep{andersson2003topup}, motion and} eddy current correction (FSL eddy) \cite{ANDERSSON20161063}, \new{and Rician noise correction \citep{koay2006rician}. \mpar{R4.4}DWIs of all time points were processed jointly using FSL eddy to avoid further coregistrations and interpolations}. Standard \new{diffusion kurtosis imaging} (DKI) weighted linear least squares fitting \cite{VERAART2013335} was applied to DWIs for calculating the diffusion and kurtosis tensors. \new{In order to compare the diffusivity time-dependence estimated based on diffusion tensor imaging (DTI) and DKI, standard DTI weighted linear least squares fitting was also applied to DWIs of b-values $\leq$ 0.4 ms/$\mu$m$^2$ for diffusion tensor calculations \citep{basser1994dti}.}\mpar{R2.2} The effective \new{$b$-value for non-diffusion weighted images}, $b^{\rm{eff}}_{0}$, included the contributions from the imaging and \new{crusher} gradients, and it was estimated to be \new{$b^{\rm{eff}}_{0}=0.001$ ms/$\mu$m$^2$} for all measured time points in this study.   

\new{To extract regions of interest (ROIs) in} gray matter, a $T_1$-weighted MPRAGE image was acquired, and the brain was segmented using FreeSurfer \new{\cite{DALE1999179,destrieux2010a2009}}. \mpar{R4.4}\new{The labels map in $T_1$-weighted image space was coregistered to the $b=0$ image space using affine transformation (FSL FLIRT) \cite{jenkinson2001flirt}, initialized with the sform/qform in the DICOM header, and was downsampled by using nearest neighbor.}  
% In addition, the computed diffusion and kurtosis parametric maps of each time point within a scan were registered to the first time point  using FSL linear registration \cite{SMITH2004S208}. 
The resulting cortical ROI is shown in Fig. \ref{Fig:Main_Fig} in red along with the $b=0$ image. 
\mpar{R1.3}
\new{To avoid white matter partial volume effects, the thresholds of fractional anisotropy FA $<$ 0.3 and $<$ 0.4 were respectively imposed to the cortical and deep gray matter ROIs based on previous studies \citep{alexander2007braindti,pfefferbaum2010deepgm}.} 
% The physical meaning of the imposed threshold lies on the fact that gray matter contains primarily cell bodies whereas white matter contains axons; this procedure delineates the cortical gray matter ROI and eliminates partial voluming. 
\new{Further, to avoid cerebrospinal fluid (CSF) signal contamination, voxels close to CSF were excluded using a CSF mask generated by FSL FAST \citep{zhang2001fast}, and was expanded by one voxel.}
Lastly, the diffusion coefficient and diffusion kurtosis for each time point and subject was calculated by averaging over all voxels of the parametric maps in the cortical gray matter ROI (Fig. \ref{Fig:Main_Fig}a and \ref{Fig:Main_Fig}c) \new{and in each gray matter sub-region (\figref{Fig:ROIs})}. 

\mpar{R2.1}\new{To compare our results with the diffusivity time-dependence observed in white matter by \citet{Fieremans2016}, white matter ROIs were also segmented by transforming John's Hopkins University DTI-based white matter atlas \citep{mori2005atlas} to the individual DWI space, as in \citep{Fieremans2016}.}

\new{
\subsection{Parameter estimation}
The three-parameter power-law relations (\ref{eq:cuDiffusion}) and (\ref{eq:cuKurtosis-inf}) were fitted to measured time-dependent mean diffusivity and kurtosis. The weighted non-linear least square fit was initialized with 1000 different combinations of initial values, and the largest cluster in parameter space was identified by using density-based clustering \citep{ester1996cluster}. We chose the median of fitted parameters within the cluster to determine the exponent $\vartheta$. 

To stabilize the three-parameter power-law fitting\mpar{R4.7}, the weight for each $t$-point was determined via Rician MRI noise propagation through DKI  pipeline, as follows: 
For one specific $t$-point,
we 
% synthesized ground-truth diffusion-weighted images with typical $D$ and $K$, 
applied Rician noise to the denoised DWIs based on the estimated noise map \citep{VERAART2016394}, 
performed DKI estimation, and repeated this procedure for 10 times to calculate the variance of estimated diffusivity and kurtosis between different noise realizations. 
% These inverse variances were used as the weights \rem{were the weights t-dependent?
% This whole par is completely unclear! Please rewrite, by first telling the reader what is your goal. Stabilize the fitting is too vague.}
The error bars for all figures was the square root of mean variance within each ROI, manifesting the noise propagation of DKI estimations. Further, we calculated weights for fitting using the inverse of mean variance within each ROI.

To evaluate the strength of the mean diffusivity and kurtosis time-dependence\mpar{R1.1}, we hypothesized that $D(t)$ and $K(t)$ are linear functions of $t^{-\vartheta}$ based on Eqs.~(\ref{eq:cuDiffusion}) and (\ref{eq:cuKurtosis-inf}) and the estimated $\vartheta$, and calculated statistical $P$-values with the null hypothesis of being no positive correlation (one-sided test). The significance level was set at 0.05 for the overall cortical gray matter, and was set at 0.002 for each gray matter sub-region (Bonferroni correction for 25 sub-regions).
}

\subsection{Structure correlation function of axonal beading}
\label{Gamma_axonsGM}

To investigate the structure of axons in gray matter \mpar{R4.5}\new{(Node 2.2.1.1 in \figref{Fig:Hierarchy})}, we processed the  data of axonal bead locations (``swellings" coinciding with synaptic boutons) in mouse cortex, originally published by \citet{Hellwig1994}. This work reports on the bead locations of 33 axons of different length, $L_m$ ($m=1...33$), ranging from approximately 100 $\mu$m to 400 $\mu$m. The construction of the power spectrum $\Gamma(k)$, Eq.~(\ref{eq:Correlator}), was performed according to following three steps:

\begin{enumerate}
\item The axonal bead density, $n(x)$, was digitized and concatenated into a single, digitized axonal line of length $L$. Note that $L\gg L_m$. 

\item The procedure of concatenation was repeated \new{200} times by randomly reshuffling the 33 axons. This procedure creates different disorder realizations. 

\item The power spectrum for each disorder realization was computed according to Eq.~(\ref{eq:Correlator}), with $V\to L$, and the Fourier transform of bead density \new{$\tilde n(k)=\int\, \d x\, e^{-ikx}n(x)$}. 
This power spectrum was then averaged over all disorder realizations. 
Note that randomly reshuffling the axons and concatenating them reduces the noise fluctuations in $\Gamma(k)$. However, after approximately \new{200} averages  the system becomes aware of the reshufflings, and averaging over subsequent reshufflings does not result in additional noise reduction in $\Gamma(k)$ \citep{PhysRevE.96.061101}. 
\end{enumerate}
% In addition, concatenating the axons into  longer segments of length $L$ results in capturing lower $k$ values in $\Gamma(k)$ as $k_{\textrm{min}}<k^m_{\textrm{min}}$ with $k_{\textrm{min}}\sim1/L$ and $k^m_{\textrm{min}}\sim1/L_m$, as long as the statistics is consistent with the short range disorder (which can be seen a posteriori, cf. Fig.~\ref{Fig:Structure} below).

% \mpar{DN re-wrote}
\subsection{MC simulations}
\label{sec:MC}

Monte Carlo (MC) simulations of Brownian motion in $d=1$ dimensions with barriers of fixed permeability $\kappa$ were performed, following a toy model of disordered axons in \figref{Fig:cartoon}a, \mpar{R4.5}\new{corresponding to the Node 2.2.1.1}. 
The ``barriers" are meant to describe, e.g., the restrictions by the narrow shafts in-between the beads (cf. Section \ref{discussion} for discussion).

The barriers were distributed in spatial dimension $d=1$ according to a PDF $P(a)$ of independent successive intervals $a$, with an average spacing between the barriers \new{$\bar{a}\simeq 4.45$ $\rm{\mu m}$} and its variance \new{$\sigma_a^2\simeq 16.4$ $\rm{\mu m^2}$}, corresponding to short-range disorder, as described by  \citet{novikov2014revealing}, Supplementary Information.
These microstructural parameters were taken to be similar to those derived from histology \cite{Hellwig1994} (as described above).

A total of five short-range disorder realizations were simulated. The barriers were distributed on a line of length \new{$L\simeq7,200$ $\rm{\mu m}$} each and approximately \new{$1,600$} barriers (restrictions) for each realization.  The number of random walkers simulated for each realization was \new{$N=1\times10^8$}. The time-step duration for each random walker was \new{$\delta t =0.002$ ms} corresponding to a spatial step size \new{$\delta x=\sqrt{2D_0\delta t}\simeq0.020\,\bar{a}$, with the intrinsic} diffusion coefficient \new{$D_0=2$ $\mu$m$^2$/ms}. \mpar{R2.3}\new{The initial positions of random walkers are randomly distributed in each realization to initialize a constant/homogeneous density.}  %how did we decide on D = 1?

We simulated membrane permeability in finite-step Monte Carlo according to \ref{sec:app-perm}. 
The probability to cross a barrier was given by Eq.~(\ref{eq:perm-prob}) with the initial barrier permeability  set to \new{$\kappa_0=0.4154$ $\mu$m/ms, such that the genuine permeability corrected for the finite time-step $\delta t$ of the simulation was $\kappa=0.4233$ $\mu$m/ms (see \ref{sec:app-perm} and \eqref{eq:kappa-correction})}. This value was chosen a posteriori to mimic the tortuosity limit 
\be \label{Dinf-SR}
D_\infty = {D_0 \over 1+\zeta} \,, \quad \zeta = {D_0 \over \kappa \bar{a}} \,, 
\ee
corresponding to \new{the membrane ``effective volume fraction" \cite{novikov2011random} $\zeta\simeq1.062$} for all MC simulations. For this model system, \citet{novikov2014revealing} found 
the coefficient 
\begin{equation} 
A = \frac{D_\infty \sqrt{\tau_r}}{\sqrt{2\pi}} \cdot \frac{\sigma_a^2}{\bar{a}^2}\left(\frac{\zeta}{1+\zeta}\right)^{3/2} , \quad c_D=2A\,, 
\label{eq:A}
\end{equation}
entering Eq.~(\ref{eq:cuDiffusion}), 
where $\tau_r = \bar a /2\kappa$ is the mean residence time within a typical interval between barriers.

The maximum diffusion time was approximately \new{$1300$ ms} corresponding to $250\tau_r$. \new{The simulated diffusivity and kurtosis were calculated based on the moments of diffusion displacements, $\langle x^2\rangle$ and $\langle x^4\rangle$.}
The random number generator used was \new{Philox4$\times$32-10 \cite{salmon2011philox}} and the MC script was developed in \new{CUDA C++. MC simulations were performed on the New York University BigPurple high-performance-computing cluster, and the total calculation time was 60 min using 5 GPU cores.}

\new{\mpar{R2.4}To evaluate the bias due to the imaging protocol and kurtosis fitting, we also simulated diffusion signals of narrow pulse with b-values = [0.1, 0.4, 1, 1.5] ms/$\mu$m$^2$ as in experiments, and fitted DKI to signals to estimate diffusivity and kurtosis.}

\subsection{Data and code availability}
All human brain MRI data for this study are available upon request. The source codes of image processing DESIGNER pipeline, power spectrum analysis, and Monte Carlo simulations can be downloaded on our github page (https://github.com/NYU-DiffusionMRI).

%%%%%%%%%%%%%%%%%%%%%%%%%%%%%%%%%%%%%
%%%%%%%%%%%%%%%%%%%%%%%%%%%%%%%%%%%%%
%%%%%%%%%%%%%%%%%%%%%%%%%%%%%%%%%%%%%
%%%%%%%%%%%%%%%%%%%%%%%%%%%%%%%%%%%%%

\section{Results}
\label{sec:results}

\subsection{$D(t)$ and $K(t)$ in human gray matter}
\label{Exp_findings}

Figs.~\ref{Fig:Main_Fig}a and~\ref{Fig:Main_Fig}c highlights the resulting axial, radial and mean diffusion coefficient and diffusion kurtosis of cortical gray matter for all subjects and time points studied in this work. The \new{noise variance $\sigma^2$} of both diffusion coefficient and kurtosis of the cortical ROI was similar for each subject and for each time point, and was approximately \new{$\sigma|_{D}\simeq 0.001$ $\rm{\mu m^2/ms}$} and \new{$\sigma|_{K}\simeq 0.005$} indicating reasonable \new{noise propagation of DKI estimation}. The observed fractional anisotropy (FA) values for the cortical ROI were approximately \new{0.18}, indicating a small anisotropy between diffusion directions; this observation allows us to focus on the mean values of the tensor diffusion metrics.

%%%%%%%%%%%%%%%%%%%%%%%%%%%%%%%%%%%%%
\begin{figure*}[th!!]
\centering
%\small
	\includegraphics[width=1.00\textwidth]{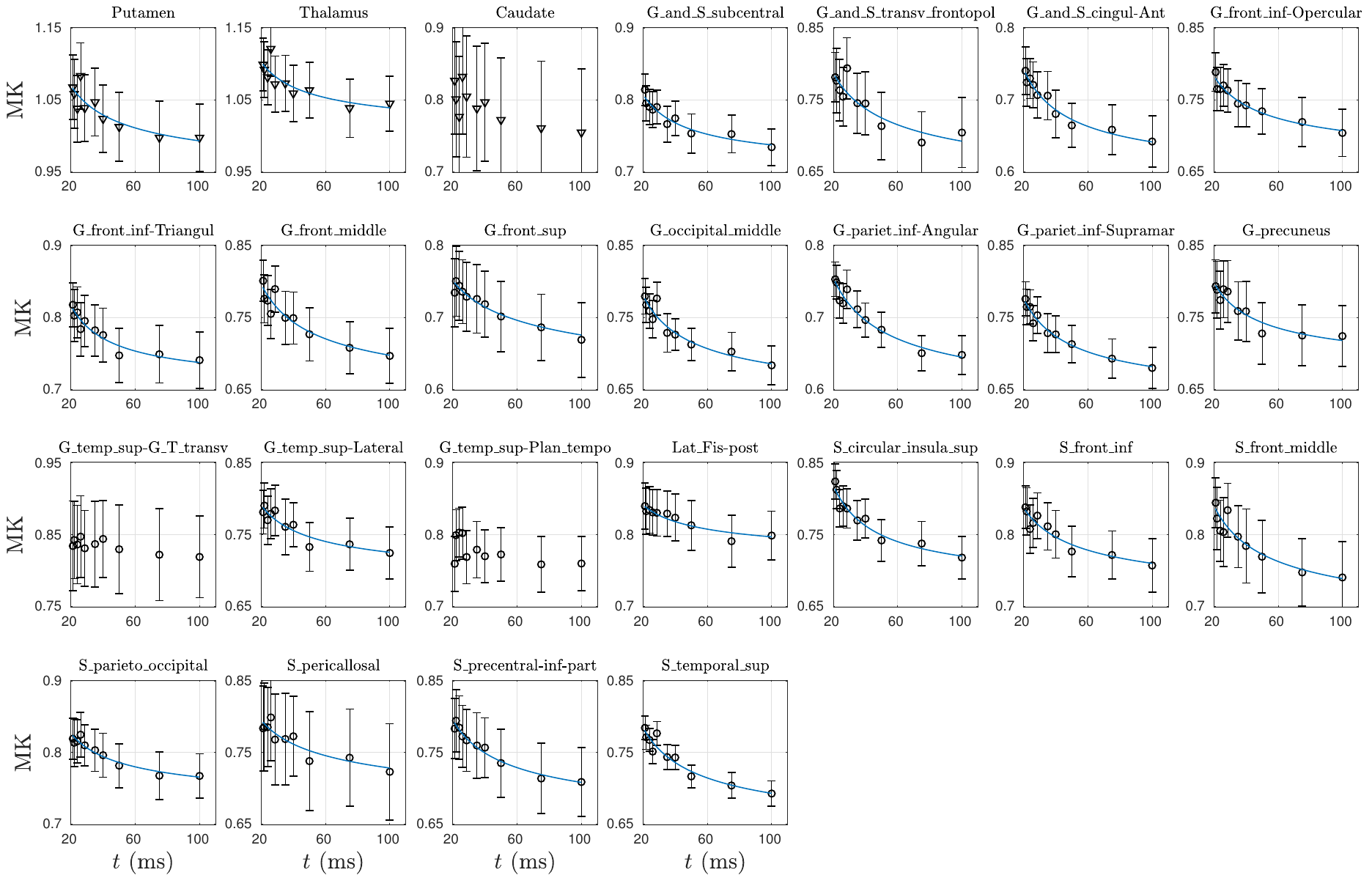}
	\caption{\textbf{Dynamical exponent for various ROIs}: \new{Significant kurtosis time-dependence averaged over 10 subjects is observed in 2 out of 3 selected ROIs in deep (triangle) gray matter and 20 out of 22 selected ROIs in cortical (circle) gray matter \citep{DALE1999179,destrieux2010a2009} (P-value $<$ 0.002), fitted with a} three degrees of freedom least squares fit according to \eqref{eq:cuKurtosis-inf}. \new{The fit parameters are in \figref{Fig:ROIs-parameters}, where the dynamical exponent $\vartheta$ is consistent among different ROIs.}}
	\label{Fig:ROIs}
\end{figure*}
%%%%%%%%%%%%%%%%%%%%%%%%%%%%%%%%%%%%%
\begin{figure*}[th!!]
\centering
%\small
	\includegraphics[width=1.00\textwidth]{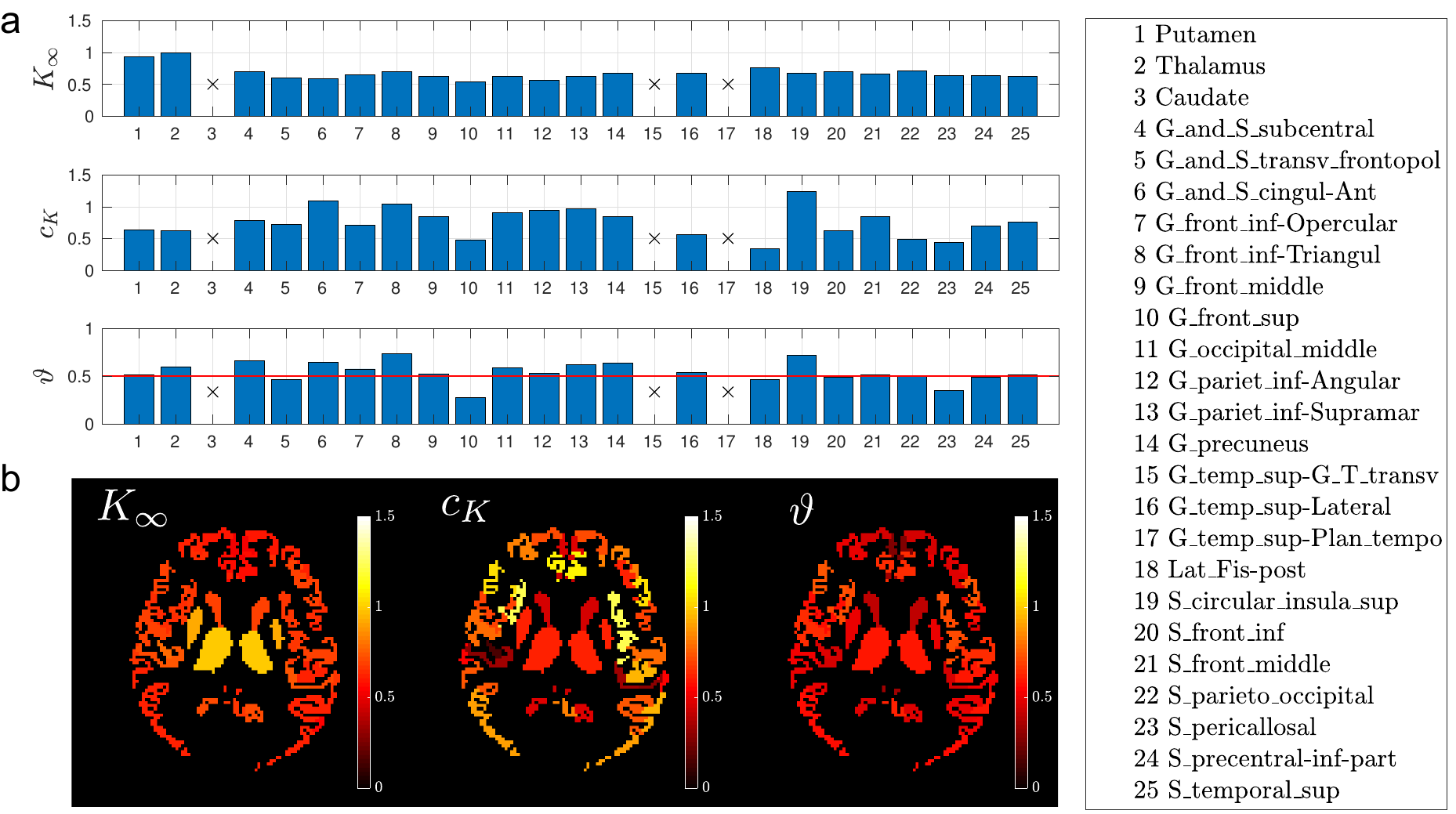}
	\caption{\new{\textbf{Fit parameters based on \eqref{eq:cuKurtosis-inf} for various ROIs}: a) Fit parameters of kurtosis time-dependence averaged over 10 subjects in ROIs of deep and cortical gray matter \citep{DALE1999179,destrieux2010a2009}. A three degrees of freedom least squares fit based on \eqref{eq:cuKurtosis-inf} leads to a dynamical exponent $\vartheta\approx 0.5$ (red line) consistent among different ROIs. The 3 ROIs showing insignificant time-dependence in kurtosis are marked by a cross. b) Fit parameters in a) are visualized in GM ROIs of a brain.}}
	\label{Fig:ROIs-parameters}
\end{figure*}
%%%%%%%%%%%%%%%%%%%%%%%%%%%%%%%%%%%%%
\begin{figure*}[th!!]
\centering
	\includegraphics[width=1.00\textwidth]{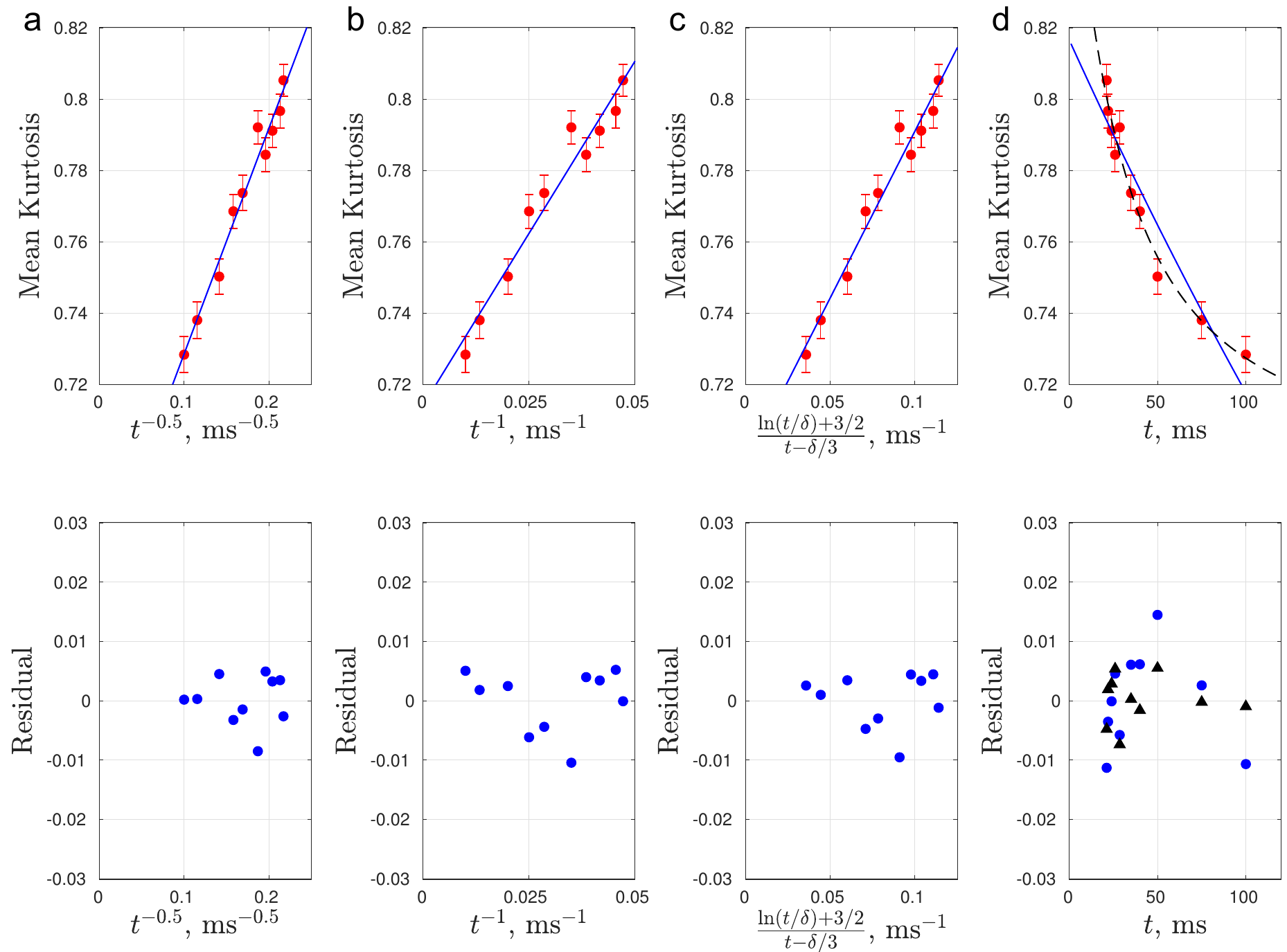}
	\caption{\textbf{Model comparison for $d=1$, $d=2$ and $d=3$ structural disorder classes, and for the K{\"a}rger model.} a) Mean kurtosis $K(t)$ in cortical gray matter plotted as a function of $t^{-0.5}$. b) $K(t)$ plotted as a function of $t^{-1}$. c) $K(t)$ plotted as a function of \new{$\left[\ln(t/\delta)+\tfrac{3}{2}\right]/(t-\delta/3)$}. d) Mean kurtosis along with the K\"arger model fit, \eqref{KurtosisKM}, with $K_\infty\equiv 0$ in blue, and the K\"arger model with an added constant $K_{\infty}$ as black dashed line. All the fit results are summarized in Section~\ref{modelFit} and \ref{sec:KMfit}. The residuals between the fit curves and measured data are shown in the bottom row.
}
	\label{Fig:Mod_Sel}
\end{figure*}
%%%%%%%%%%%%%%%%%%%%%%%%%%%%%%%%%%%%%
%%%%%%%%%%%%%%%%%%%%%%%%%%%%%%%%%%%%%
	\begin{figure*}[th!!]
\centering
	\includegraphics[width=1.00\textwidth]{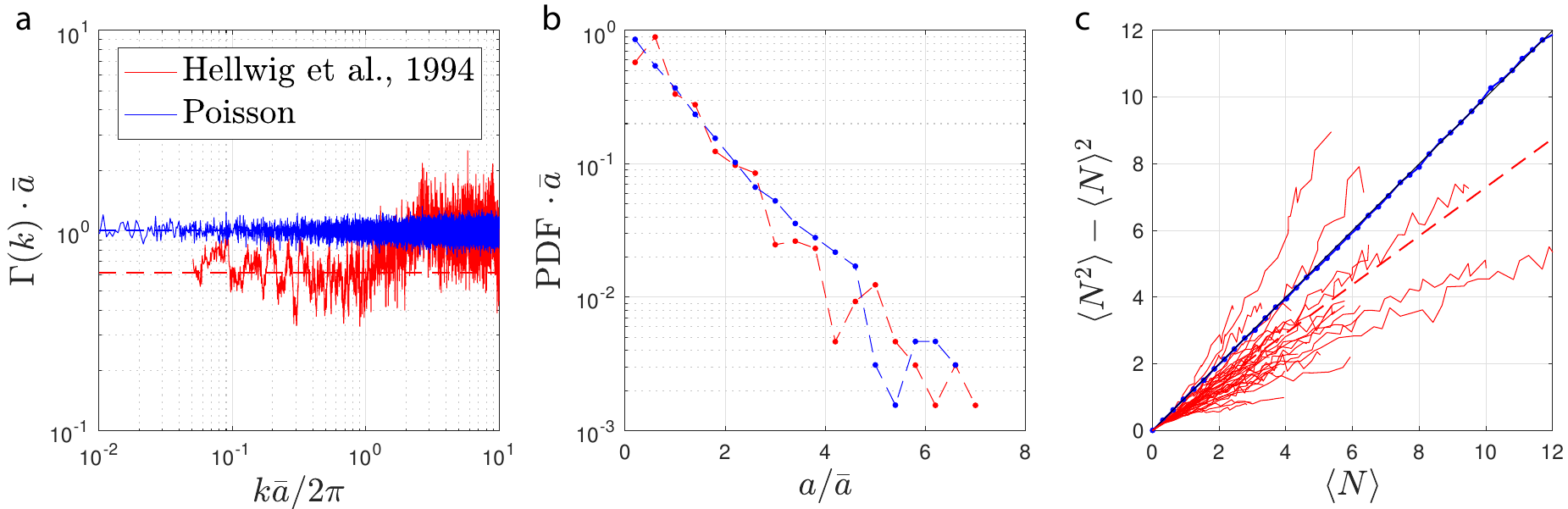}
	\caption[]{\textbf{Short-range disorder is revealed in cortical gray matter.} \mnote{R4.9}
% 	{\bf a,} A power-law of $\vartheta=0.52$ is revealed in the temporal scaling of mean kurtosis in cortical gray matter, consistent with the expected power-law for short-range disorder of $\vartheta=1/2$.
	a) \new{Power spectrum}, $\Gamma(k)\cdot\bar{a}$ calculated via Eq.~(\ref{eq:Correlator}), of axonal beadings in the cortex based on \citep{Hellwig1994} (red) shows a plateau lower than unity as $k\rightarrow0$. \new{Power spectrum} for the strictly Poissonian disorder is also shown for comparison (blue), with a unity plateau as $k\rightarrow0$.  \new{b) The corresponding histogram of bead distance in cortex (red) and strictly Poissonian disorder (blue).} c) Scaling of number of axonal beadings within a varying window with respect to the variance for each of thirty-three axons taken from \cite{Hellwig1994}. The dashed red line indicates a linear fit of all the red lines with a slope of 0.73, which diverges from the unity line corresponding to the simulated Poissonian disorder (blue).}
	\label{Fig:Structure}
\end{figure*}
%%%%%%%%%%%%%%%%%%%%%%%%%%%%%%%%%%%%%

By performing an average of the mean diffusivity and kurtosis over all subjects, a distinct time-dependence was observed in the diffusion kurtosis at the time scale of the experiment as shown in \figref{Fig:Main_Fig}d, \news{cf. \figref{fig:suppl-MK} for mean kurtosis histogram of each subject}. On the other hand, the diffusivity showed \new{relatively weak} time-dependence (\figref{Fig:Main_Fig}b). 
\mpars{R2.6}\news{This is also indicated by the normalized diffusion signals plotted versus diffusion weighting $b$ and diffusion time $t$ in cortical gray matter (\figref{fig:suppl-S-b}): At low $b<1$ ms/$\mu$m$^2$, diffusion signals do not obviously increase or decrease with diffusion time; in contrast, at $b=$ 1.5 ms/$\mu$m$^2$, diffusion signals visibly decrease with diffusion time. Therefore, the corresponding diffusivity time-dependence (estimated mainly from low $b$ data) is weak and nosiy, whereas the observed kurtosis time-dependence (estimated mainly from high $b$ data) is more significant.}
\mpar{R1.2\\R2.1}\new{Based on the tissue length scale in histology \citep{Hellwig1994,GlantzLewis} and MC simulations in Section \ref{sec:result-mc}, diffusion in cortical gray matter is in the long time regime for $t> 20\,$ms, allowing us to probe the structural disorder, Node 2.2.1, by studying the dynamical exponent $\vartheta$.}

The comparison of DTI and DKI results showed that, while DKI yielded slightly larger diffusivity values than DTI in most of the brain ROIs, the diffusivity time-dependence given by DTI and DKI was nearly identical \mpars{R2.5}\news{(\figref{fig:suppl-AD-MD} and \figref{fig:suppl-RD})}. Furthermore, in brain white matter ROIs, we also observed significant axial and radial diffusivity time-dependence in this dataset, consistent with the previous study \citep{Fieremans2016}.
% which may point towards diffusion kurtosis being more sensitive to subtle features in the structure. This difference in the scaling between the two metrics is reflected in the coefficients of Eq.~(\ref{c'-theta}).   

\subsection{Estimation of dynamical exponent $\vartheta$ (Node 2.2.1)} % can you refer here to the matching Node in FIg 1?
\label{modelFit}

Eq.~(\ref{eq:cuKurtosis-inf}) was used to estimate the observed dynamical exponent from the subject-averaged mean kurtosis. The result was \new{$\vartheta =0.56$} after performing a three degrees of freedom least squares fit, with \new{$c_K=0.70$ and $K_{\infty}=0.68$, and $\chi^2/\rm{DOF}=1.04$}. 

\figref{Fig:Scaling_subjects_statistics}a shows the mean kurtosis for all subjects scanned in this work along with statistics of the three degrees of freedom parameter fit to Eq.~(\ref{eq:cuKurtosis-inf}). Relatively high $\chi^2/\rm{DOF}$ was observed for each fit in comparison to the global, as shown in \figref{Fig:Scaling_subjects_statistics}b. On the other hand, reasonable agreement was observed between the fitted values $\vartheta$, $c_K$ and $K_{\infty}$ of each subject (\figref{Fig:Scaling_subjects_statistics}c-d-e). 
% Also Fig. \ref{Fig:Structure}a shows a dynamical exponent of $\vartheta=0.52$ which results from a two degrees of freedom least squares fit using $K_{\infty}$ from Fig. \ref{Fig:Scaling_subjects_statistics}.

 \figref{Fig:ROIs} highlights the scaling of mean kurtosis \new{for 25 additional ROIs of sub-regions in deep and cortical gray matter}, in comparison with the global cortical gray matter. A reasonable agreement was observed between the global dynamical exponent \new{$\vartheta=0.56$} and $\vartheta$ for each ROI \new{in \figref{Fig:ROIs-parameters}}.

\figref{Fig:Mod_Sel} shows the measured mean kurtosis of the global cortex with respect to $t^{-0.5}$, $t^{-1}$, \new{$\left[\ln(t/\delta)+3/2\right]/(t-\delta/3)$}, and $t$. A straight line was observed in Figure \ref{Fig:Mod_Sel}a-b when kurtosis is plotted with respect to both $t^{-0.5}$ and $t^{-1}$,  with \new{$\chi^2/\rm{DOF}\sim0.9$ and 1.4 respectively}. In addition, a straight line was observed in \figref{Fig:Mod_Sel}c when kurtosis is plotted with respect to \new{$\left[\ln(t/\delta)+3/2\right]/(t-\delta/3)$} \citep{Burcaw2015} with \new{$\chi^2/\rm{DOF}\sim1.1$}. This observation reveals that the fit does not allow for a statistically confident model selection between the three functional forms. 

\new{\mpar{R1.1}Instead, we can select models in Node 2.2.1 by comparing the time-dependence of diffusivity and kurtosis, i.e., the ratio $\xi$ of $c_K$ to ($c_D/D_\infty$) in \eqref{xi}. In \figref{Fig:Mod_Sel}a and \ref{Fig:Mod_Sel}c, the ratio $\xi=2.43$ and $2.41$ for $t^{-0.5}$ and $\left[\ln(t/\delta)+3/2\right]/(t-\delta/3)$ power-law indicates that the short-range disorder in $1d$ ($\xi=2$) is the most preferred model in Node 2.2.1.}
We discuss these findings further in  Section \ref{discussion}.
% \rem{with other compartments contributing, this ratio will not be as clean as theory predicts, since $K$ are not  adding up. Comment on this? Maybe estimate the error?}

%\rem{
%\subsection{Any section on $(\ln t)/t$ fits?}
%Can the exponent $0.61$ be the ``average" between $t^{-1/2}$ in the neurites and $(\ln t) /t$, which is nearly $1/t$ ? 
%This is kind of the obvious thing to assume; also, note that $c_K$ has large numerical factor, $6$ or $42/5=8.4$ --- could it be the reason we see $K(t)$ but not $D(t)$?
%}

\subsection{K{\"a}rger model's parameter estimation (Node 1.1.2)} % can you refer here to the matching Node in FIg 1?
\label{sec:KMfit}

If we were instead to adopt the exchange picture between Gaussian compartments, fitting the KM kurtosis (\ref{KurtosisKM}) to the observed mean kurtosis would yield an exchange time between compartments, which are most likely to be the 
 neurites (dendrites and axons) and the extra-neurite space. \figref{Fig:Mod_Sel}d shows the measured mean kurtosis and the fit of Eq.~(\ref{KurtosisKM})  (with the added constant $K_{\infty}$) to the data in black dashed line. The fit had a \new{$\chi^2/\rm{DOF}\simeq 0.97$ and an} exchange time value \new{$\tau_{\rm{ex}}^{\rm KM}\simeq11$ ms}. 
%that is about 2 orders of magnitude lower than all previous estimates. 
 On the other hand, a fit of the original K{\"a}rger Model (setting $K_{\infty}\equiv0$) yields an exchange time of \new{$\tau_{\rm{ex}}^{\rm KM}\simeq250$ ms with a relatively poor $\chi^2/\rm{DOF}\simeq 3.2$. The above estimated exchange times, either with or without a non-zero $K_\infty$, are out of the range of our measurements with diffusion time $t=21.2-100$ ms.} 

%%%%%%%%%%%%%%%%%%%%%%%%%%%%%%%%%%%%%%%%%%%%%%%%%%%%%%
\subsection{Structure correlation function of axons in gray matter}
\label{structure}

%\rem{Overall criticism: reads disconnected from Sec.~\ref{Gamma_axonsGM}}

%%%%%%%%%%%%%%%%%%%%%%%%%%%%%%%%%%%%%
\begin{figure*}[th!!]
\centering
	\includegraphics[width=1.0\textwidth]{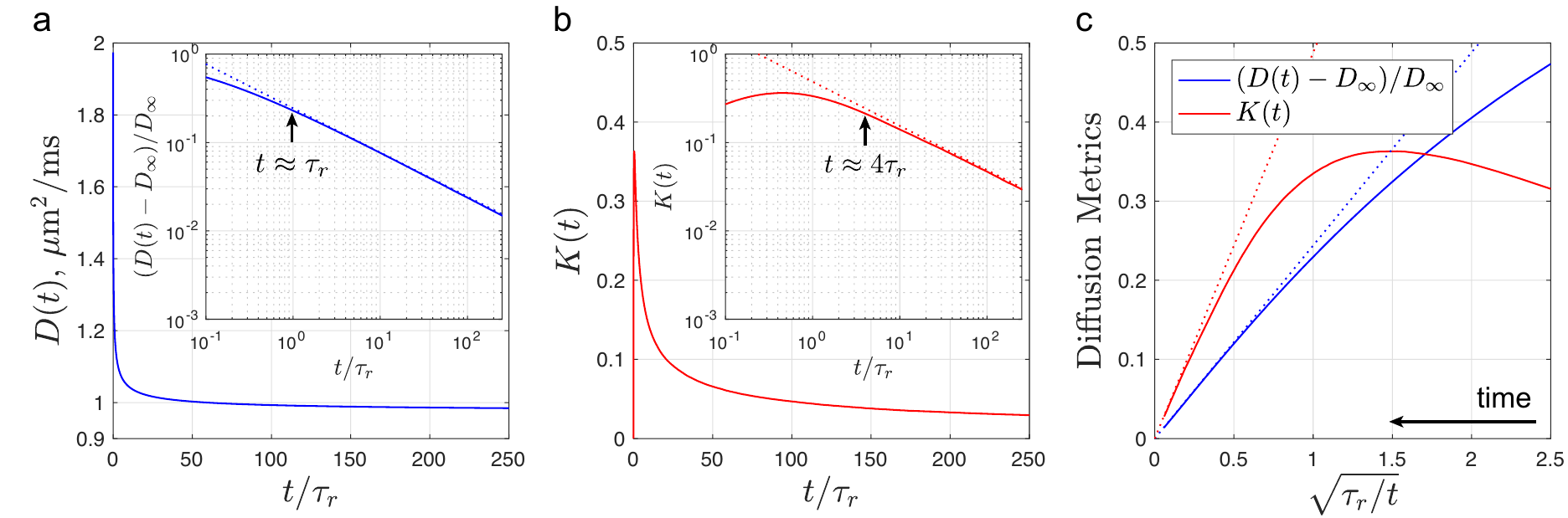}
	\caption{\textbf{Simulated diffusivity and kurtosis on a one-dimensional short-ranged disorder line along with theoretical predictions Eqs.~(\ref{c'-theta}) and (\ref{Dinf-SR})--(\ref{eq:A}). Diffusivity and kurtosis reveal that the long-time limit starts at \new{$t\sim\tau_r$ and $t\sim4\tau_r$ respectively}.} a) Simulated diffusivity with respect to \new{$t/\tau_r$} approaches its tortuosity value $D_{\infty}$. Inset: Normalized diffusivity along with the theory (dashed line) given by Eqs.~(\ref{c'-theta}) and (\ref{Dinf-SR})--(\ref{eq:A}). b) Similarly, the simulated kurtosis with respect to \new{$t/\tau_r$} approaches zero at long times but shows a non-monotonic behavior at \new{$t\lesssim\tau_r$}. Inset: Short-time limit of diffusion kurtosis shows an initial increase and a plateau at \new{$t\lesssim\tau_r$} before reaching the long-time limit for $t>\tau_r$ agreeing with the theory (dashed line) given by Eqs.~(\ref{c'-theta}) and (\ref{Dinf-SR})--(\ref{eq:A}). c) Simulated diffusivity and kurtosis (solid lines) plotted with respect to $\sqrt{\tau_r/t}$ approach a straight line for \new{$t\gg \tau_r$} where the long-time limit empirically starts. The two-fold difference in the coefficients $c_K$ and $c_D/\Dinf$ from Eq.~(\ref{c'-theta}) is apparent in the two-fold difference in the slopes of the simulated quantities for \new{$t\gg \tau_r$}.}
	\label{Fig:simulations}
\end{figure*}
%%%%%%%%%%%%%%%%%%%%%%%%%%%%%%%%%%%%%

%According to the above results, \new{the observed time-dependence of diffusivity and kurtosis in gray matter is most probably caused by the short-range disorder in 1d (Node 2.2.1.1). Here}, by investigating the microstructure of gray matter neurites, we could \new{further explore} the origins of the time-dependence in gray matter.
%
%%it looks like the effects of the exchange are negligible, and the value of $\vartheta$ is sufficiently close to the $\vartheta=1/2$ for the one-dimensional short range disorder along the neurites, previously identified by \citet{novikov2014revealing} in the data of \citet{Does2003}. While the ``flat" $D(t)$ seems puzzling, here we adopt the $d=1$ short range disorder as a working hypothesis (at the very least consistent with the observed $K(t)$), and explore its possible anatomical origins using Monte Carlo simulations. 
%
%Beads and spines along gray matter neurites hinder diffusion and the underlying statistics of their position must have an effect on the low-$k$ scaling \new{in its power spectrum} $\Gamma(k)$ determining the structural exponent $p$ according to Eq.~(\ref{eq:Correlator}). \mpar{R3.3} %\citet{Hellwig1994} measured the beads' positions in 33 axons of mouse cerebral cortex and concluded, based on a statistical analysis, that the bead occurrence along axons follows Poissonian statistics (ideally Poissonian would have $\sigma=\bar a$) i.e. fully uncorrelated bead placements, with $\bar{a}\simeq4.5\,\mu$m (mean distance between beads) and a variance of $\sigma^2_a\simeq 16\,\mu$m$^2$. 

\new{We now study the low-$k$ behavior of the power spectrum  \eqref{eq:Correlator} of bead placement density $n(x_0)$ quantified from the measurements by \citet{Hellwig1994} in mouse cerebral cortex, to determine the structural exponent. 
It is useful to consider the dimensionless  $ \Gamma(k)\cdot \bar a$, which is equal to unity 
for Poissonian statistics, as shown  in \figref{Fig:Structure}a (blue) for simulated fully uncorrelated barrier placement.} 

For general short-range disorder, residual correlations give rise to a plateau in $\Gamma(k)|_{k\to 0}\cdot \bar a$ different from unity. \new{Based on histology}, a plateau of approximately 0.6 was observed after constructing the structure correlation function (\ref{eq:Correlator}) shown in \figref{Fig:Structure}a in red for the bead placements of \cite{Hellwig1994}. 
This indicates that bead occurrence along axons corresponds to a short-range disorder, confirming the structural exponent $p=0$ announced in Node 2.2.1.1, Section~\ref{theory}. In addition, \figref{Fig:Structure}b highlights the PDF $P(a)$ of the successive intervals for artificially constructed Poissonian disorder, and for the experimentally measured axonal bead placements from \cite{Hellwig1994}. Although noisy, a maximum of the PDF for the bead placement (in red) at \new{$a/\bar{a}\approx 0.6$} may distinguish it from the perfectly exponential PDF $=(1/\bar a)\cdot e^{-a/\bar a}$ (linear in semi-logarithmic scale) for the Poissonian statistics (blue). % this is a very long, can you split in two or three?

An alternative approach for distinguishing Poissonian statistics is investigating the scaling of the mean number of restrictions $\langle N\rangle$ within a window of length \new{$L_N$} with respect to their variance $\langle N^2\rangle - \langle N\rangle^2$ within this window \cite{Shepherd6340}.
Fig.~\ref{Fig:Structure}c shows such scaling. 
% of the mean number of restrictions (beads) within the window of varying length with respect to its variance within the same window. 
For Poissonian statistics,  $\langle N^2\rangle - \langle N\rangle^2=\langle N\rangle$ is expected, as shown by the blue line. On the other hand, the solid red lines represent this scaling for the 33 individual axons measured in \cite{Hellwig1994} along with a fit over all axons shown by the dashed line.
As expected, for short-range disorder, $\langle N\rangle$ grows in proportion to $\langle N^2\rangle - \langle N\rangle^2$ but with a  slope \new{$\approx0.73$} different from $1$. 
% This slope $\approx 0.7$ matches well the plateau  $\Gamma(k)|_{k\to0}\cdot \bar a = (\sigma/\bar a)^2 \approx 0.6$ (cf. ref.~\cite{novikov2014revealing}, Supplementary Eq.~S13 for the derivation). 

% \new{
% The above $d=1$ picture was first introduced by \citet{novikov2014revealing} to reveal and interpret the  $\omega^{\vartheta}$ scaling of the oscillating gradient diffusion measurement of \cite{Does2003} in rat cortical gray matter, for which $\vartheta=1/2$ was found. It is remarkable that the same power law exponent $\vartheta=1/2$ is here observed in human cortical gray matter. 
% Together with our direct quantification of mouse cortical structural disorder from \cite{Hellwig1994}, this suggests that 
% \begin{enumerate}[(i)]
% \item the $p=0$ short-range disorder in one dimension is a universal microstructural signature of structural heterogeneity in neurites across mammals; and
% \item it manifests itself in the $t$-dependent dMRI signal acquired over macroscopic voxels in vivo, and hence, can be quantified and monitored in disease, development and aging. 
% \end{enumerate}
% }

What is short-range disorder qualitatively, and why is it ubiquitous? The hallmark of short-range disorder is the finite correlation length $l_c$, beyond which the correlation function \mpar{R4.1}\new{$\langle n(\x_0+\x)n(\x_0) \rangle_{\x_0}$} decays sufficiently fast (this applies in any dimension, not just in $d=1$), so that the ``memory" about where one should  expect another restriction is forgotten for \new{$x \gg l_c$}. In other words, for such large \new{$x$}, one could view \new{the correlation function
$\sim \delta(\x)$ as} a $\delta$-function of the width $\sim l_c$. Hence, in the $k$-space, the power spectrum of such a localized object is approximately constant, $\Gamma(k)\sim k^0 = \const$ for all $k \lesssim 1/l_c$, 
yielding the structural exponent $p=0$. 
%In our one-dimensional example, this yields exponent $\vartheta=1/2$ according to Eq.~(\ref{vartheta}). 

We note that other alternatives for the placement of the restrictions are the {\it hyperuniform disorder}, $p>0$, with ``almost-periodic" placements of the restrictions (characterized by the suppressed structural fluctuations at large distances) that emerge, e.g., due to effective repulsion of restrictions, or can be artificially created  \cite{PhysRevE.96.061101}; and the so-called {\it strong disorder}, with $p<0$, such that the power spectrum (\ref{eq:Correlator}) diverges at $k\to 0$ \cite{novikov2014revealing}.

%\rem{this sentence is unclear:} 
%As this is an alternative approach for distinguishing between short range and Poissonian statistics, investigating the low-$k$ scaling of $\Gamma(k)$ is preferred as it captures the long-range fluctuations which manifest themselves in the scaling of diffusion and kurtosis.
 %%%%%%%%%%%%%%%%%%%%%%%%%%%%%%%%%%%%%
%%%%%%%%%%%%%%%%%%%%%%%%%%%%%%%%%%%%%
%%%%%%%%%%%%%%%%%%%%%%%%%%%%%%%%%%%%%
%%%%%%%%%%%%%%%%%%%%%%%%%%%%%%%%%%%%%
%%%%%%%%%%%%%%%%%%%%%%%%%%%%%%%%%%%%%

\subsection{MC simulations in $d=1$: Diffusion metrics}
\label{sec:result-mc}
%\rem{reads disconnected from Sec.~\ref{sec:MC}. Clearly state the goals: see if we can test Eq.~(\ref{c'-theta}); see if we observe overall less pronounced variation of $D(t)$ vs that in $K(t)$}

To investigate the sensitivity of the diffusion coefficient and diffusion kurtosis to the microstructural features and validate Eq.~(\ref{c'-theta}), we performed Monte Carlo simulations in dimension $d=1$. \figref{Fig:simulations}a-b highlighted the time-dependence of the diffusion coefficient and diffusion kurtosis up to $t=250\tau_r$, and both metrics reached the tortuosity limit already by that time. Diffusivity approached the tortuosity limit, \new{$D_{\infty}=0.97\,\mu$m$^2$/ms}, for times $t\gg\tau_r$, where diffusion become effectively Gaussian. Similarly, $K(t)$ approached zero for $t\gg\tau_r$. In addition, the kurtosis showed a non monotonic time-dependence at approximately \new{$t\lesssim \tau_r$} where a maximum was observed as shown in the inset of Fig. \ref{Fig:simulations}b. 

%Since diffusion takes place in a short-range disordered transmission line with $d=1$ and structural exponent $p=0$, the expected dynamical exponent, with which diffusivity and kurtosis would reach their tortuosity limit, is $\vartheta=1/2$. 
\figref{Fig:simulations}c shows the simulated diffusivity and kurtosis as a function of $\sqrt{\tau_r/t}$, so that a straight line is formed at long times according to Eq.~(\ref{c'-theta}). Good agreement was observed between Eq.~(\ref{c'-theta}) and the simulated diffusivity and kurtosis at long times. It is observed that the system is in the long-time limit at already \new{$t\approx \tau_r$ for diffusivity and $t\approx 4\tau_r$ for kurtosis (insets of \figref{Fig:simulations}a-b)}, which effectively means that the molecules then already have traversed a couple of mean barrier spacings $\bar{a}$. 

In addition, \figref{Fig:simulations}c highlights the simulated diffusivity and kurtosis along with \new{the theoretical prediction Eqs.~(\ref{Dinf-SR})--(\ref{eq:A}) for the slope $2A/D_\infty$ and $4A/D_\infty$ (dotted lines)}. The scaling of the diffusion kurtosis for long times reveals that the system is at the long time limit at approximately \new{$t\gtrsim4\tau_r$} which may point to both diffusivity and kurtosis being equally robust metrics. However, kurtosis $t$-dependence is observed to be relatively twice more pronounced than that of the diffusivity tail $(D(t)-\Dinf)/\Dinf$ due to the two-fold difference in the coefficients following from Eq.~(\ref{c'-theta}). In simulations, at approximately \new{$t\gtrsim 4\tau_r$}, the tails in diffusivity and kurtosis are indeed observed to differ by a factor of $2$ (\figref{Fig:simulations}c).

It is not unexpected that in cortical gray matter and for the shortest diffusion time $t>20\,$ms studied in this work, the diffusion is effectively in the long time limit, since spines are placed in dendrites with mean spacing \new{$\bar{a}\simeq3-3.4$ $\mu$m} \cite{GlantzLewis}, and beads \new{are placed in axon collaterals with $\bar{a}\simeq 2.4-7.5$ $\mu$m} \cite{Hellwig1994}. Another important observation extracted from MC simulations is that kurtosis may be the more sensitive metric for observing subtle effects such as time-dependence in cortical gray matter.
% , an observation that was also revealed in our experimental data of Fig. \ref{Fig:Main_Fig}.
% \rem{your data does not reveal this. It shows no t-dependence in $D$. You may only speculate that's one of the reasons, but you cannot ignore testing the $\vartheta=1$ models.}

\new{\mpar{R2.4}Further, the simulation of signals compared to moments revealed that DKI fitting yields a small bias in diffusivity ($<0.1$\% bias in $D_\infty$ and 3\% in $c_D$) and a moderate bias in kurtosis (14\% bias in $c_K$), with the same $t^{-0.5}$ functional form valid at the same time scale (data not shown).}

\section{Discussion}
\label{discussion}

%\rem{the whole Discussion needs rewriting, given earlier suggestions. I give guidelines along the way but you have to think it through first, depending on what you are actually selling}

In this study, we provided \new{experimental} evidence of time-dependent kurtosis in human gray matter at time-scales from $21.2-100$ ms, whereas diffusivity showed \new{relatively weak and noisy} time-dependence during the same time scales. \new{Here}, we discuss the interpretation of the observed time-dependence in diffusion kurtosis based on the scenarios introduced in Section~\ref{theory}, and connect them with the underlying microstructure of neurites in gray matter.

Diffusion in gray matter intra-axonal space may be hindered by spines and beads along dendrites and axons which occur at length scales of approximately $3-6\,\mu$m \news{\citep{Hellwig1994,GlantzLewis}}. Diffusion along the neurites may be modeled as that along one dimensional structurally-disordered channels as shown in Fig. \ref{Fig:cartoon}a. \mpars{R2.1}\news{The corresponding correlation time along neurites is $t_c\sim l_c^2/(2dD_0)\sim$ $2-9$ ms for $D_0=2\,\mu$m$^2$/ms and $d=1$. This correlation time scale is much shorter than the applied diffusion times ($>$ 20 ms), and thus the diffusion along neurites in this study falls into the long time regime.} In this scenario, which corresponds to Node 2.2.1.1 of Fig. \ref{Fig:Hierarchy} (cf. Section \ref{theory}), the time-dependent kurtosis should scale with a power-law of $\vartheta=1/2$. A three degrees of freedom fit of \eqref{eq:cuKurtosis-inf} to the experimental data revealed a power-law of \new{$\vartheta=0.56$} with a reasonable \new{$\chi^2/\text{DOF}=1.04$}, which is sufficiently close to the theoretical $\vartheta=1/2$. Therefore, \new{with negligible water exchange between intra- and extra-neurite spaces and low extra-neurite volume fraction (cf. 20\% in adult rat cortex \cite{Bondareff1968}), }
% in the presence of no exchange between intra-extra neurite space, and since the volume fraction of extra neurite space is approximately 20\% \cite{Bondareff1968} in adult rat cortex, and a similar fraction should be expected in humans, 
the measured signal may be originating primarily from the intra-neurite space (at least in areas of gray matter that are not dominated by the cell bodies), pointing towards 1{\it d} short-range disorder. 

% An additional evidence pointing towards this scenario is the statistics of 
\new{This scenario is also consistent with histology when analyzing the} axonal bead placement in gray matter. The class of disorder, i.e., the statistics of restriction placement, may have an effect on the observed time-dependence. Short-range disorder is generally ubiquitous in Physics and in Biology, hence the structural exponent $p=0$ is not unexpected. The short-range character of restriction placement is supported by the experimental data of $\Gamma(k)$ at low-$k$ values, shown in Fig. \ref{Fig:Structure}, suggesting that beads along axons are distributed according to a PDF with a finite mean and variance according to short-range disorder ($p=0$). \mpar{R1.6} \new{In addition to the beads, \citet{morales2014dendrite} also observed that dendritic spines in adult human neocortex are mostly randomly positioned, further supporting the character of short-range disorder in gray matter.}

The observed value of diffusivity at the tortousity limit of Fig. \ref{Fig:Main_Fig} is approximately \new{$D_{\infty}\simeq 0.97$ $\mu$m$^2$/ms}, which allows us to provide an estimate of the permeability of the one-dimensional ``barriers" (e.g., shafts between neurite beads) after mapping their complex structure onto a $d=1$ dimensional transmission line of barriers of permeability $\kappa$. As mentioned earlier, $\bar{a}\simeq 3\,\mu$m for the spines and beads along neurites, which combined with \new{$D_{\infty}\simeq 0.97\,\mu$m$^2$/ms (this study) and $D_0\simeq 2\,\mu$m$^2$/ms \citep{rotinv}}, results in \new{$\zeta\simeq 1.06$ and $\kappa\simeq0.63\,\mu$m/ms based on \eqref{Dinf-SR}}.

\new{
The above 1{\it d} picture was first introduced by \citet{novikov2014revealing} to reveal and interpret the  $\omega^{\vartheta}$ scaling of the oscillating-gradient diffusion measurement of \cite{Does2003} in rat cortical gray matter, for which $\vartheta=1/2$ was found. It is remarkable that the same power law exponent $\vartheta=1/2$ is here observed in human cortical gray matter. 
Together with our direct quantification of mouse cortical structural disorder from \cite{Hellwig1994}, this suggests that 
\begin{enumerate}[(i)]
\item the $p=0$ short-range disorder in one dimension is a universal microstructural signature of structural heterogeneity in neurites across mammals; and
\item it manifests itself in the $t$-dependent dMRI signal acquired over macroscopic voxels in vivo, and hence, can be quantified and monitored in disease, development and aging. 
\end{enumerate}
}

\new{Another possible scenario is hindered diffusion in the} extra-neurite space, which is abundant with cells, ions and metabolic substrates \cite{Nicholson1981}. \new{Extra-neurite space can} be modeled as a two- or three-dimensional\mpar{R3.4\\R4.10} random medium (depending on its anisotropy), such that the diffusion is restricted either transverse to a fiber bundle (two-dimensional geometry, cf. \figref{Fig:cartoon}b), or  in $3d$ by the randomly placed ``rods" (cf. \figref{Fig:cartoon}c). As discussed in Section \ref{theory}, the microstructure of each compartment and the dimensions will have an effect on the observed time-dependence of the diffusion metrics. Diffusion in the extra-neurite space would yield a power-law exponent of  $\vartheta=1$ with a logarithmic singularity in the kurtosis \new{$K(t)\sim \ln( t/\tilde{t}_c)/t$} in the case of $d=2$ and $p=0$ (short-range disorder) as well as $d=3$ and $p=-1$ (extended disorder). In cases where $p>0$, the kurtosis would scale as $\sim 1/t$ at long times. Plotting the experimental data with respect to $1/t$ and \new{$\ln( t/\tilde{t}_c)/t$} did not reveal any important features that may point to one scenario or the other as the least squares fits were equally reliable (see Fig. \ref{Fig:Mod_Sel} and section \ref{Exp_findings}). \mpar{R1.1}\new{However, the ratio $\xi\simeq 2.4$ between the tails in $K(t)$ and $D(t)$, in both \figref{Fig:Mod_Sel}a and \ref{Fig:Mod_Sel}c, cf. \eqref{xi}, is much closer to $\xi=2$ than to $\xi=6$ or $42/5$. This ratio further indicates that 1{\it d} short-range disorder ($\vartheta=1/2$ and $\xi=2$), originating from the intra-neurite space, is the most preferred model in Node 2.2.1. The estimated ratio $\xi$ is not exactly 2 probably due to contributions of diffusivity and kurtosis time-dependence in other compartments, e.g., extra-neurite space and astrocytes, with different (and non-dominant) power-law exponents. To sum up, for the first time, the comparison of diffusivity and kurtosis time-dependence ($c_D/D_\infty$ and $c_K$) reveals the structural disorder in tissue micro-geometry.} 

\new{The last scenario to be discussed is that of exchange, here approximated by the K{\"a}rger Model. KM with a non-zero $K_\infty$ yields an exchange time $\tau_{\rm{ex}}^{\rm KM}\simeq11\,$ms ($\chi^2/\rm{DOF}=0.97$), contradicting the underlying assumption of slow exchange regime \citep{KM}. 
Furthermore, a fit of the original KM (setting $K_\infty\equiv0$) yields an exchange time $\tau_\text{ex}^\text{KM}\simeq$ 250 ms with a relatively poor fit quality ($\chi^2/\text{DOF}=3.2$). Both exchange time estimates, using KM with or without $K_\infty$, are out of the range of our measurements ($t=$ 21.2-100 ms), hence their reliability cannot be high. \mpar{R1.1}More importantly, significant diffusivity time-dependence was observed in cortical gray matter, inconsistent with an expected time-independent diffusivity in KM. 
\mpars{R2.2}\news{Unlike the structural disorder power-law scaling in Node 2.2.1, KM is based on an assumption of exchange between two Gaussian diffusion pools and thus cannot explain the diffusivity time-dependence, a signature of non-Gaussian diffusion in at least one of the compartments.}
Hence, we conclude that KM and related exchange cannot be used to explain the observed diffusivity and kurtosis time-dependence. 

\mpars{R2.2}\news{Further, for both power-law scaling and KM, the overall $K(t)$ should approach zero as $t\to\infty$. However, within a voxel, there are always some partial-volume contributions from tissues of different diffusivities (e.g., CSF), without the possibility of exchange for observable diffusion times. This macroscopic heterogeneity leads to a constant kurtosis, which we denote as $K_\infty$ and add to $K(t)$ expression to account for all such partial-volume contributions.}

At the same time, however, we cannot exclude a possible contribution of exchange (as a physical effect, beyond a relatively primitive KM) to our observed data. While exchange time $\tau_\text{ex}>1000\,$ms was in vivo measured in lenticular nucleus and thalamus using FEXI \citep{Lampinen2017}, and $\tau_\text{ex}\simeq$ 115 ms was found between extra-neurite space and astrocytes in vitro \citep{Donghan2014}, much shorter exchange time range $\tau_\text{ex}\simeq$ 10-30 ms was recently found in human gray matter on a human Connectome scanner in the high-$b$ regime, at $b\lesssim 25\,\units{ms/\mu m^2}$ \citep{veraart2018gm}.
Furthermore, exchange times of about $10\,$ms were found in live and fixed excised neonatal mouse spinal cord between membrane structures and free environments using DEXSY \citep{williamson2019spine}.}
\newt{Therefore, we speculate that human gray matter may be in the {\it crossover regime}, where the exchange effects compete with those of the structural disorder (Node 2.2.2); in this picture, exchange is likely to affect the numerical coefficients, such as $D_\infty$, $c_D$ and $c_K$, whereas the qualitative $t^{-1/2}$ power-law scaling is determined by the structural disorder.}
\mpart{R2.1}\newt{This prompts the generalization of the present exchange K\"arger model (Node 1.1.2) by incorporating non-Gaussian diffusion properties in each compartment (Node 2.2.2). We can capture the complexity of this theoretical question through the approximate signal representation of a special 1{\it d} case, given by \citet{grebenkov2014theory}.
}

A few experimental limitations may not allow us to extract more accurate exchange times and parameters of the scaling laws from the data. First, a larger diffusion time window is necessary in order to accurately campture the power-law dependence of Eqs.~(\ref{eq:cuDiffusion}) and (\ref{eq:cuKurtosis-inf}), as well as to fit the K{\"a}rger Model. While a $T_1$-weighted sequence of the type of STEAM allows for longer diffusion times, it may also introduce artificial time-dependence in the diffusivity and kurtosis due to molecular exchange \new{between compartments (e.g., myelin water and intra-/extra-cellular water in white matter)} during the STEAM storage times \cite{HHLee}. \new{To rule out the latter confounding factor, we used spin-echo sequence with fixed TE and TR to fix the $T_1$-weighting and exchange effect between compartments.} In addition, a smaller voxel size may be beneficial in order to allow for a more accurate ROI selection and better statistics in the estimation of the diffusion coefficient and kurtosis. \mpar{R2.5}\new{A possible approach to simultaneously evaluate water exchange and structural disorder (Node 2.2.2) is to extend the effective medium theory to other more advanced sequences/diffusion gradient waveforms \cite{JespersenISMRM2019}, such as the FEXI sequence \citep{lasic2011fexi}.}

\section{Conclusions}
\label{Conc}

%\rem{rethink this based on analysis for $\vartheta=1$, etc}
%In this work, time-dependent kurtosis was observed for the first time in human gray matter at time scales from 21.2--100\,ms. The observed time-dependence was modeled in three ways. First, as non-Gaussian diffusion along intra-neurite space, and mapped into one-dimensional transmission line similar to Fig. \ref{Fig:cartoon}a. Within this one-dimensional approach, a detailed investigation of the statistics of bead occurrence was performed and showed that axons are most likely short-range disordered with a structural exponent of $p=0$. Second, we considered non-Gaussian diffusion in a higher dimensional  extra-neurite space. And last, we considered Gaussian diffusion yet with the presence of exchange between intra- and extra-neurite space, using a K{\"a}rger model approach. Our experimental results favor \new{the first} scenario, \new{and} the microstructural analysis of bead density in axons points to diffusion in intra-neurite space. On the other hand, diffusion in human gray matter at these time scales may be in the crossover regime where both the intra-compartmental microstructure of intra- and extra-neurite space, as well as the exchange between them, play a similarly important role.

\new{
In this work, time-dependent kurtosis was observed for the first time in human gray matter at time scales  $t=21.2-100\,$ms. 
Using the proposed model selection tree of Fig.~\ref{Fig:Hierarchy} for time-dependent diffusivity and kurtosis, we conclude that 1{\it d} structural disorder along the one-dimensional neurites plays the dominant role. The estimated dynamical exponent $\vartheta\approx 1/2$ suggests that diffusion along neurites is affected by short-range disorder (randomly positioned restrictions), consistent with histological results, and the observed power-law is different from that of the K\"arger model (\ref{KurtosisKM}), $K(t)\sim1/t$ in long time limit ($t\gg\tau_\text{ex}$). Furthermore, the exchange time ($\simeq 11$ ms) given by K\"arger model is out of our measurement range, as well as contradicting the KM assumption of slow exchange regime. Therefore, the observed time-dependence occurs due to physics beyond the KM. Exchange may contribute to the observed $D(t)$ and $K(t)$, such that the diffusion in human gray matter at these time scales may be in the crossover regime, where the exchange competes with the structural disorder (Node 2.2.2 in selection tree), while the disorder sets the overall $t^{-1/2}$ scaling of $D(t)$ and $K(t)$. In conclusion, while model-selection is not fully resolved, we present a compelling case of the sensitivity of time-dependent dMRI to the structural disorder along the neurites in the gray matter.}

\section{Acknowledgements}

We would like to thank Jelle Veraart, Benjamin Ades-Aron and Gregory Lemberskiy for fruitful discussions throughout the entire process of data acquisition, data processing and manuscript writing,  
\new{Thorsten Feiweier for developing advanced diffusion WIP sequence, 
and BigPurple High Performance Computing Center of New York University Langone Health for numerical computations on the cluster. Research was supported by the National Institute of Neurological Disorders and Stroke of the NIH under awards R01 NS088040 and R21 NS081230, and was performed at the Center of Advanced Imaging Innovation and Research (CAI2R, www.cai2r.net), an NIBIB Biomedical Technology Resource Center (NIH P41 EB017183).

\figref{Fig:cartoon}a is adapted with permission from \news{\citep{woolley1990dendrite}, Copyright 1990 Society for Neuroscience}, and \citep{Shepherd6340}, \news{Copyright 2002 National Academy of Sciences}.
}

%%%%%%%%%%%%%%%%%%%%%%%%%%%%%%%%%%%%%
%%%%%%%%%%%%%%%%%%%%%%%%%%%%%%%%%%%%%
%%%%%%%%%%%%%%%%%%%%%%%%%%%%%%%%%%%%%
%%%%%%%%%%%%%%%%%%%%%%%%%%%%%%%%%%%%%
%%%%%%%%%%%%%%%%%%%%%%%%%%%%%%%%%%%%%

\appendix
\section{Relation between power-law tails in $D(t)$ and $K(t)$}
\renewcommand{\theequation}{A.\arabic{equation}}
\setcounter{equation}{0}
\label{sec:app-power}

\subsection{Power-law tails for a single compartment}
Diffusion in the long time limit ($t\gg t_c$) effectively homogenizes a sample's microstructure \cite{novikov2014revealing}, mapping the problem onto that characterized by a smoothly varying local diffusivity \new{$D(\x_0)$} with a mean $\overline{D}$ and a small variation \new{$\delta D(\x_0)=D(\x_0)-\overline{D}$} relative to the mean \citep{novikov2010emt}. 
The crucial observation is that the spatial fluctuations of \new{$D(\x_0)$} mimic those of the microstructural restrictions \new{$n(\x_0)$} at large \new{displacement $\x$}; in particular, the power spectrum 
\be \label{GammaD}
\Gamma_D(k) = {D(-\k) D(\k)\over V} \simeq B \cdot k^p \,, \quad k\to 0
\ee
is characterized by the same structural exponent $p$ as in Eq.~(\ref{eq:Correlator}). 

In what follows, we will relate the power-law tails in $D(t)$ and $K(t)$ to the effective medium parameter $B$ of Eq.~(\ref{GammaD}), based on the perturbative treatment up to the order ${\cal O}\lp\delta D^2\rp$, i.e., up to the first order in the power spectrum (\ref{GammaD}). Our starting point is the cumulants  of molecular displacements, 
given by Eqs. (24)--(25) of \citep{novikov2010emt}: 
\bea \label{eq:x2cumulant}
\langle x^2 \rangle &=& 2! \int \frac{{\rm d}\omega}{2\pi}e^{-i\omega t} \frac{{\cal D}(\omega)}{(-i \omega_+)^2}\,,
\\ \label{eq:x4cumulant}
\langle x^4 \rangle &=& 4! \int \frac{{\rm d}\omega}{2\pi} e^{-i\omega t} \left[ \frac{\Sigma_4(\omega)}{(-i\omega_+)^2} + \frac{{\cal D}^2(\omega)}{(-i\omega_+)^3} \right]\,,
\eea
with $\Sigma_4(\omega)$ explained later. The symbol $\omega_+$ denotes that the integration is calculated on a complex plane of $\omega$, and all poles are in the lower half-plane as a result of causality, cf. Appendix A of \citep{review-nbm}.

The dispersive diffusivity in Eqs.~(\ref{eq:x2cumulant})--(\ref{eq:x4cumulant}) is given by Eq.~(7) of \citep{novikov2014revealing}
\be \label{Dw}
\D(\w) - \Dinf = {-i\w \over d \Dinf} \int\! {\d^d\k\over (2\pi)^d}\, {\Gamma_D(k)\over -i\w + \Dinf k^2}
\ee
equivalent to the instantaneous diffusion coefficient (\ref{eq:instDiffusion})
\begin{equation} \label{Dinst} 
D_{\rm inst}(t) - D_\infty \simeq \frac{1}{dD_\infty}\int\! \frac{\d^d {\bf k}}{(2\pi)^d}\, \Gamma_D(k)\, e^{-D_\infty k^2 t} 
= A \cdot t^{-\vartheta}
\end{equation}
where, for $\Gamma_D(k)$ from Eq.~(\ref{GammaD}), we obtain 
\begin{equation} \label{eq:c-B}
A = \frac{B \cdot \Omega_d \cdot \Gamma_{\rm E}(\vartheta)}{2d \cdot (2\pi)^d \cdot D_\infty^{1+\vartheta}} \,.
\end{equation}
Here, $\Omega_d = 2\pi^{d/2}/\Gamma_E(d/2)$ is the surface area of a unit sphere in $d$ dimensions ($\Omega_d = 1,2\pi,4\pi$ for $d = 1,2,3$), and $\Gamma_{\rm E}(\cdot)$ is Euler's $\Gamma$-function.
Using either Eq.~(\ref{Dw}) or the relation
${\cal D}(\omega) = -i\omega \int e^{i\omega t} D_{\rm inst}(t)\, \d t$  (cf. Section~2.2.2 of \cite{review-nbm}), 
we find in the frequency domain 
\begin{equation} \label{eq:dispersive-diff}
{\cal D}(\omega) \simeq D_\infty + A\,   \Gamma_{\rm E}(1-\vartheta) \cdot (-i\omega)^\vartheta\,.
\end{equation} 
Hence, using Eq.~(\ref{D=Dinst}), we find 
\be \label{x2}
\langle x^2 \rangle = 2D(t) \, t \simeq 2\Dinf t + 2c_D \cdot t^{1-\vartheta} \,, 
\ee
and $c_D$ is defined in Eq.~(\ref{eq:cuDiffusion}).

As we can see from Eq.~(\ref{eq:x4cumulant}), the dispersive diffusivity alone is not enough to calculate the kurtosis.  We will now show that, in general, the fourth order dispersive kinetic coefficient, Eq. (33) of \cite{novikov2010emt}, 
\begin{equation} \label{S4}
\begin{split}
\Sigma_4(\omega) = &\int\! \frac{{\rm d}^d {\bf k}}{(2\pi)^d}\, \Gamma_D(k)\left[ G^0_{\omega,{\bf k}} - \frac{5}{d}\, \overline{D}k^2\left(G^0_{\omega,{\bf k}}\right)^2 \right.\\
&\left. + 4 \beta_d \, (\overline{D}k^2)^2 \left( G^0_{\omega,{\bf k}} \right)^3 \right],
\end{split} \end{equation}
originating from expanding the self-energy part up to $q^4$, gives a  contribution to the scaling of the 4th-order moment that is of the same order of magnitude as the second term in Eq.~(\ref{eq:x4cumulant}). 
The angular average 
\be \label{beta}
\beta_d = \langle \cos^4\theta \rangle = \frac{\int_0^\pi\! \d\theta \, \sin^{d-2}\theta\, \cos^4\theta}{\int_0^\pi\! \d\theta \, \sin^{d-2}\theta}
= {3\over d(d+2)} 
\ee
 in the spherical coordinates in $d$ dimensions, entering the last term of Eq.~(\ref{S4}), can be expressed  using the spherical volume element by reducing the integrals to Euler's $B$-functions. As $\Sigma_4$ is already small in the perturbation theory parameter 
 $\Gamma_D$, we can substitute $\overline{D} \to \Dinf$ (tortuosity asymptote) there, as well as in the free propagator in the $\w,\q$ representation
\begin{equation}
G^0 _{\omega,{\bf q}} = \frac{1}{-i\omega + D_\infty q^2} \,.
\end{equation}

Using the power spectrum (\ref{GammaD}) in Eq.~(\ref{S4}) and reducing all the integrals to Euler's $B$-functions by the substitution 
$y= \Dinf k^2/(-i\w)$, after straightforward algebra we obtain 
\begin{equation} \label{eq:sigma4}
\Sigma_4(\omega) = \frac{p(3p+d-4)}{2(d+2)} \Gamma_{\rm E}(1-\vartheta)  \cdot A \Dinf  (-i\omega)^{\vartheta-1} 
\end{equation}
with $A$ given by Eq.~(\ref{eq:c-B}). We can see that, e.g., for the short-range disorder in any dimension, $p=0$, the contribution 
$\Sigma_4(\w)$ vanishes, but in general it does not --- e.g., for the case of $p=-1$ in $d=3$ considered in Node 2.2.1.2 of Fig.~\ref{Fig:Hierarchy}. 

%For 1-dimensional case, using $\Gamma_D(k) \propto k^p$ at low-$k$ (eq. (\ref{eq:Correlator})), we obtain
%\begin{equation*}
%\Sigma_4(\omega) \propto p(1-p) \cdot \frac{i}{8\overline{D}} \left( e^{i\frac{\pi}{4}}  \sqrt{\frac{\omega}{\overline{D}}} \right) ^ {p-1} \,,
%\end{equation*}
%showing that $\Sigma_4(\omega)$ does not contribute to kurtosis when $p=0$ (short-range disorder) corresponding to $\vartheta=1/2$ for $d=1$ dimensions.

%Similarly, for 2-dimensional case, we have
%\begin{equation*}
%\Sigma_4(\omega) \propto p(2 - 3p) \cdot \frac{i}{64\overline{D}} \left( e^{i\frac{\pi}{4}}  \sqrt{\frac{\omega}{\overline{D}}} \right) ^p \,,
%\end{equation*}
%indicating that $\Sigma_4(\omega)$ has no contribution to kurtosis if $p=0$ (short-range disorder), corresponding to $\vartheta=1$ for $d=2$ dimensions.

We are now ready to calculate $\langle x^4 \rangle$ by substituting Eqs.~(\ref{eq:sigma4})  and (\ref{eq:dispersive-diff}) into Eq.~(\ref{eq:x4cumulant}). We perform the integration in the complex plane of $\omega$ by rotating the path of integration to pass along the two sides of the branch cut of $\w^{\vartheta}$ which is convenient to choose along the negative imaginary axis. 
In this way, we obtain 
\begin{equation} \label{eq:x4-emt}
\langle x^4 \rangle = 12D_\infty^2 t^2 + 24 \left( 2 + \frac{p(3p+d-4)}{2(d+2)} \right) \cdot \frac{A D_\infty\, t^{2-\vartheta}}{(2-\vartheta)(1-\vartheta)}.
\end{equation}
%\begin{equation} \label{eq:x4-emt}
%\langle x^4 \rangle = 12 D_\infty^2 t^2 -48 A D_\infty \left( \frac{1}{2-\vartheta} - \frac{1}{1-\vartheta} \right)t^{2-\vartheta}\,.
%\end{equation}
The leading term of $\langle x^2 \rangle^2$ (to the order $\O(\delta D^2) \sim \O(A)$), using Eq.~(\ref{x2}), reads
\begin{equation} \label{eq:x2sq-emt}
\langle x^2 \rangle^2 \simeq 4 D_\infty^2t^2 + 8 c_D D_\infty \, t^{2-\vartheta} \,.
\end{equation}
Finally, using  the definition $K(t) \equiv \langle x^4 \rangle/\langle x^2 \rangle ^2 -3$, and again keeping only the lowest-order terms in $c_D\sim A$ (cf. Eq.~(\ref{eq:cuDiffusion})), we obtain our main analytical result --- Eq.~(\ref{eq:cuKurtosis}) with 
\begin{equation} \label{eq:kurtosis-emt}
c_K =  \frac{6 c_D }{D_\infty } \cdot \left[ \left( 2 + \frac{p(3p+d-4)}{2(d+2)} \right) \cdot \frac{1}{2-\vartheta} - 1 \right] ,
%K(t) \simeq \frac{6 c \cdot t^{-\vartheta}}{D_\infty (1-\vartheta)} \cdot \left[ \left( 2 + \frac{p(3p+d-4)}{2(d+2)} \right) \cdot \frac{1}{2-\vartheta} - 1 \right]. 
\end{equation}
yielding the ratio (\ref{xi}) in the main text. 
While the scaling with $A/\Dinf$ of the result (\ref{eq:kurtosis-emt}) could be guessed from the dimensional considerations, the dependence on $p$ and $d$ is nontrivial. Remarkably, due to the $\Sigma_4$ term, the tail in kurtosis depends {\it separately} on $p$ and $d$, rather than  on the exponent (\ref{vartheta}) alone. Therefore, measuring the tails in both $D(t)$ and $K(t)$ can allow one to determine the structural exponent $p$ and the effective dimensionality $d$ separately, whereas measuring just the diffusion coefficient only yields their sum. 
%\begin{equation} \label{eq:kurtosis-emt}
%K(t)\equiv \frac{\langle x^4 \rangle}{\langle x^2 \rangle ^2} -3 \simeq \frac{6\vartheta}{(2-\vartheta)(1-\vartheta)} \frac{A}{D_\infty} t^{-\vartheta} \,,
%\end{equation}
%which is only applicable to short-range disorder in 1- and 2-dimensions since $\Sigma_4(\omega)$ may not be zero for other cases.

%The short-range disorder in 1-dimension ($p=0, d=1$) has a dynamical exponent $\vartheta=\frac{1}{2}$. Substituting into eq. (\ref{eq:kurtosis-emt}), the time-dependent kurtosis is given by
%\begin{equation*}
%K(t) \simeq \frac{4c}{D_\infty} \frac{1}{\sqrt{t}}\,,
%\end{equation*}
%which is equivalent to eq. (\ref{c'-theta}). Note that for short-range disorder $c=\frac{D_{\infty} \sqrt{\tau_r}}{\sqrt{2\pi}}\frac{\sigma^2}{\bar{a}^2}\left(\frac{\zeta}{1+\zeta}\right)^{3/2}$. 

While obtaining Eq.~(\ref{c'-theta}) from Eq.~(\ref{eq:kurtosis-emt}) for $p=0$ and $d=1$ is straightforward, the $\vartheta=1$ case is formally singular. To resolve this singularity in Eq.~(\ref{eq:kurtosis-emt}), we take an $\epsilon\to0^{-}$ limit of $\vartheta = 1 + \epsilon$ in Eq.~(\ref{eq:cuKurtosis}). 
For instance, for $p=0$ and $d=2$, 
\begin{equation*}
K(t) \simeq \frac{6}{-\epsilon} \frac{A}{D_\infty} \frac{t^{-\epsilon}}{t}\,,
\end{equation*}
where $t^{-\epsilon} = \exp(-\epsilon \ln t) \simeq 1-\epsilon \ln t $, leading to
\begin{equation} \label{eq:kurtosis-epsilon}
K(t) \simeq \frac{6A}{D_\infty} \frac{ \ln t - \frac{1}{\epsilon} }{t} \,.
\end{equation}
To better understand the physical meaning of the regularizer $1/\epsilon$ (reminiscent of the dimensional regularization in quantum field theory), we explore a similar singularity in the dispersive diffusivity in Eq.~(\ref{eq:dispersive-diff}):
\begin{equation} \label{eq:dispersive-diff-epsilon}
{\cal D}(\omega) \simeq D_\infty + A \cdot i\omega \left[ \ln(-i\omega) + \frac{1}{\epsilon} \right]\,,
\end{equation}
where we used the Laurent expansion of the Euler's $\Gamma$-function $\Gamma_{\rm E}(-1-\epsilon) \simeq \frac{1}{\epsilon}$ and $(-i\omega)^ \epsilon \simeq 1+\epsilon \ln(-i\omega)$ to simplify this formula. The singularity in Eq.~(\ref{eq:dispersive-diff-epsilon}) originates from the time scale $\sim \tilde{t_c}$ from which the power-law tail begins. \citet{Burcaw2015} showed that, for $p=0$ and $d=2$, the dispersive diffusivity is given by 
\begin{equation} \label{eq:dispersive-diff-tc}
{\cal D}(\omega) \simeq D_\infty + A \cdot i\omega \ln(-i\omega \tilde{t_c})\,.
\end{equation}
Comparing Eqs. (\ref{eq:dispersive-diff-epsilon}) and (\ref{eq:dispersive-diff-tc}), we  identify $\frac{1}{\epsilon}$ with $\ln \tilde{t_c}$, which after substituting into Eq.~(\ref{eq:kurtosis-epsilon}) yields Eq.~(\ref{eq:instKurtosisSR2}).

Similar considerations yield  Eq.~(\ref{eq:instKurtosisExtend}), albeit with a coefficient that has a nontrivial contribution from $\Sigma_4(\w)$ due to nonzero $p$. 

\new{
We note that the  reason for the singularities at $\vartheta=1$ is the fact that we measure the cumulative, rather than the instantaneous diffusion coefficient, for which the integral in Eq.~(\ref{D=Dinst}) becomes insensitive to the tails decreasing faster than $1/t$. This is a feature of our PGSE measurement, rather than of the underlying physics of diffusion. Similar considerations apply to the (cumulative) kurtosis. Had we worked with the instantaneous 2nd and 4th order cumulants, e.g., defining them via $\partial_t \langle x^n(t)\rangle_c$, such a problem would not have arisen. The tail ratio (\ref{xi}) therefore can be generalized onto the power-law tails of the suitably defined instantaneous diffusivity and kurtosis for any $\vartheta$. 
}
%\subsection*{Extended disorder in 3-dimension}
%Extended disorder in 3-dimensions ($p=-1, d=3$), such as randomly oriented rods, has a dynamical exponent $\vartheta = 1$. Assuming that $\vartheta = 1 + \epsilon$, $|\epsilon|\ll 1$, and substituting into eq. (\ref{eq:kurtosis-emt}), we have
%\begin{equation*}
%K(t) \simeq \frac{-42}{5\epsilon} \frac{A}{D_\infty} \frac{t^{-\epsilon}}{t}\,.
%\end{equation*}
%Similarly, using $t^{-\epsilon}\simeq1-\epsilon \ln t$ and $\frac{1}{\epsilon}\simeq \ln \tilde{t_c}$, we obtain eq. (\ref{eq:instKurtosisExtend}).

\news{
\subsection{Eqs.~(\ref{xi}) and (\ref{eq:cuKurtosis-inf}) for multiple compartments}
}
For Node 2.2.1, if multiple compartments are present, their power-law tails compete, and the ones with the slowest tail (smallest $\vartheta$) dominate at long enough $t$, as it has been the case for the time-dependent diffusion \cite{novikov2014revealing}. 
Hence, without the loss of generality,  it suffices to consider the case when all compartments have the same (the smallest) exponent $\vartheta$, differing only by their individual effective-medium parameters $c_{i,D}$, $c_{i,K}$ and bulk diffusivities $D_{i,\infty}$. 
For compartments with no time-dependence or with larger $\vartheta$ values, one can set $c_{i,D}$ and $c_{i,K}$ to zero. 

Generalizing \eqref{K0}, the overall kurtosis of multiple {\it non-Gaussian} compartments is given by \citep{Dhital2017}
\begin{equation} \label{eq:K-general}
    \overline{K}=3\frac{\text{var}(D_i)}{\langle D_i\rangle^2} + \frac{\langle D_i^2\cdot K_i\rangle}{\langle D_i\rangle^2}\,,
\end{equation}
with the $i$-th compartment's diffusivity $D_i(t)$ and kurtosis $K_i(t)$. 
% In this case, the overall diffusivity and overall kurtosis, based on Eqs.~(\ref{D_non_interacting}), (\ref{K0}), (\ref{eq:cuDiffusion}) and (\ref{eq:cuKurtosis}), are given by
Substituting Eqs.~(\ref{eq:cuDiffusion}) and (\ref{eq:cuKurtosis}) into Eqs.~(\ref{D_non_interacting}) and (\ref{eq:K-general}) yields
\be \notag
\overline{D}(t) \simeq \overline{D}_\infty + \overline{c}_D\cdot t^{-\vartheta}\,,\quad
\overline{K}(t) \simeq \overline{K}_\infty + \overline{c}_K\cdot t^{-\vartheta}\,,
\ee
where
\be \notag
\overline{D}_\infty=\langle D_{i,\infty} \rangle\,,\quad
\overline{c}_D = \langle c_{i,D}\rangle\,,
\ee
we obtain
\be \notag
\overline{K}_\infty = 3\frac{\text{var}(D_{i,\infty})}{\overline{D}_\infty^2}\,,\quad
\ee
\be \label{eq:K-overall}
\overline{c}_K = \frac{1}{\overline{D}_\infty^2}\left[\langle D_{i,\infty}^2\cdot c_{i,K}\rangle + 6\left(\langle D_{i,\infty} \cdot c_{i,D}\rangle - \frac{\overline{c}_D}{\overline{D}_\infty}\langle D_{i,\infty}^2\rangle \right) \right]\,.
\ee
% with the {\it i}-th compartment's time-dependent parameters ($c_{i,D}$, $c_{i,K}$) and bulk diffusivity ($D_{i,\infty}$). 
Further, substituting the definition of $\xi(p,d)$ in \eqref{xi} for the single compartment into \eqref{eq:K-overall}, we can represent $\overline{c}_K$ as
\be \notag
\overline{c}_K= \frac{1}{\overline{D}_\infty^2}\left[(\xi+6)\cdot \langle D_{i,\infty}\cdot c_{i,D} \rangle - 6\frac{\overline{c}_D}{\overline{D}_\infty}\langle D_{i,\infty}^2\rangle\right]\,.
\ee
As a result, for the whole system, the dimensionless ratio $\overline{\xi}$ of the tails $\overline{K}(t)$ and $[\overline{D}(t)-\overline{D}_\infty]/\overline{D}_\infty$ is given by
\be \notag
\overline{\xi}\equiv \frac{\overline{c}_K}{\overline{c}_D/\overline{D}_\infty}=(\xi+6)\cdot\frac{\langle D_{i,\infty} \cdot c_{i,D}\rangle}{\langle D_{i,\infty}\rangle \langle c_{i,D}\rangle} - 6 \frac{\langle D_{i,\infty}^2\rangle}{\langle D_{i,\infty}\rangle^2}\,,
\ee
which indicates that \eqref{xi} is applicable to multiple compartments, i.e., $\overline{\xi}=\xi$, if all $D_{i,\infty}$ are equal to each other. \\
% any one of the following conditions is satisfied:
% \begin{enumerate}
%     \item The $D_{i,\infty}$ is the same for all compartments.
%     \item The $D_{i,\infty}$ and $c_{i,\infty}$ are linearly dependent, i.e. $D_{i,\infty}/c_{i,\infty}=$ const is the same for all compartments.
% \end{enumerate}
%     \item The $c_{i,\infty}$ is the same for all compartments.
%     \item The $D_{i,\infty}$ and $c_{i,\infty}$ are independent, i.e. the covariance $\text{cov}(D_{i,\infty},c_{i,\infty})\equiv \langle D_{i,\infty}\cdot c_{i,\infty}\rangle-\langle D_{i,\infty}\rangle \langle c_{i,\infty}\rangle=0$. This is the most general condition covering also the previous two.
% \end{enumerate}
% Based on the covariance defined above, we further conclude that $\overline{\xi}>\xi$ if $\text{cov}(D_{i,\infty},c_{i,\infty})>0$ and vice versa.

\new{
\section{Accurate simulation of membrane permeability for finite-step MC simulations}\label{sec:app-perm}}
\renewcommand{\theequation}{B.\arabic{equation}}
\setcounter{equation}{0}
\renewcommand{\thefigure}{B.\arabic{figure}}
\setcounter{figure}{0}

% \subsection{The physics of the permeability correction. Equal molecular concentrations}

The  permeation probability $P$ through a membrane of permeability $\kappa$ depends on the distance $\delta s$ between the random walker and the encountered membrane when the distance is smaller than the step size $\delta x=\sqrt{2dD_0\delta t}$, with $D_0$ the intrinsic diffusivity and $\delta t$ the time-step in $d$ dimensional space, as derived in Appendix A of \citep{KM}, Eq.~(43):
\begin{equation} \label{P-Els}
    \frac{P}{1-P}=\frac{2\kappa\delta s}{D_0}\,.
\end{equation}
The functional form of $P$ is well-regularized even for the highly permeable membrane: the limit $\kappa\to\infty$ yields probability $P\to1$, as expected.

However, calculating the distance from random walkers to encountered membranes can be slow in actual implementations, especially for simulations using complicated shapes. To simplify simulations, we would like to approximate $\delta s$ with $\delta x$, 
by {\it averaging} over possible step sizes $\delta s$ (up to $\delta x$), and introducing a constant factor $C_d$ into Eq.~(\ref{P-Els}) to account for this approximation in $d$ dimensions. For that, let us first assume  low probability ($P\ll1$), such that the denominator in the left-hand side of Eq.~(\ref{P-Els}) can be neglected. This yields the permeation probability
\begin{equation} \label{eq:perm-prob}
    P\simeq \frac{\kappa_0 \delta x}{D_0}\cdot C_d\,,
\end{equation}
where $\kappa_0$ is the input permeability value (whose difference from the genuine $\kappa$ will be explained below), 
and $C_d$ = 1, $\pi$/4, and 2/3 for $d$ = 1, 2, and 3 \cite{powles1992simulation,szafer1995simulation,FIEREMANS201839}.  \eqref{eq:perm-prob} is applicable when the assumption $P\ll1$ is satisfied, i.e.,
\begin{equation*}
    \kappa_0\ll\sqrt{\frac{D_0}{2d\delta t}}\cdot\frac{1}{C_d}\,,
\end{equation*}
indicating that the larger $\kappa_0$, the smaller the time step $\delta t$ needs to be; in this case, the resulting $\kappa\approx \kappa_0$.

Let us now extend this approximation of $\delta s$ by $\delta x$ onto large $\kappa$, for which the input $\kappa_0$ would be significantly different from $\kappa$. It turns out that averaging over $\delta s$ simply renormalizes the input $\kappa_0$ entering Eq.~(\ref{eq:perm-prob}), to achieve a genuine $\kappa$, Eq.~(\ref{eq:kappa-correction}) below. 
To derive this result, one needs to realize that averaging over $\delta s$ influences not only the permeation probability but also the calculation of particle flux density. To solve this problem, we demand the Fick's first law to be satistifed with the permeation probability \eqref{eq:perm-prob} and derive a correction factor that renormalizes the permeability $\kappa_0 \to \kappa$. 

Considering a permeable membrane with an input $\kappa_0$ for the calculation of $P$, the intrinsic diffusivity over the left and right sides of the membrane is $D_1$ and $D_2$ respectively. 
Approximating the particle density variation close to the membrane with a linear function of the distance from the membrane, we obtain (details to be published elsewhere) the genuine membrane permeability
\begin{equation} \label{eq:kappa-correction}
    \kappa = \frac{\kappa_0}{1-\alpha}\,,
\end{equation}
where
\begin{align} \notag %\label{eq:kappa-factor}
    \alpha &= \frac{1}{2}\kappa_0\left( \frac{\delta x_1}{D_1} + \frac{\delta x_2}{D_2}\right)\cdot C_d\\ \label{eq:kappa-factor-prob}
    &=\frac{P_{1\to2} + P_{2\to1}}{2}\,,
\end{align}
and $P_{1\to2}$ is the permeation probability from left to right given by \eqref{eq:perm-prob}.

For example, in Section \ref{sec:MC}, we applied $\delta t=$ 0.002 ms, $D_1=D_2=$ 2 $\mu$m$^2$/ms, and $\kappa_0=$ 0.4154 $\mu$m/ms in $1d$ simulations, yielding  $\alpha=$ 0.019 and a 2\% correction to the actual permeability $\kappa=$ 0.4233 $\mu$m/ms.

\bigskip
%%%%%%%%%%%%%%%%%%%%%%%%%%%%%%%%%%%%
% \newpage
\bibliographystyle{elsarticle-harv}
\bibliography{gray_matter}
%\bibliographystyle{ieeetr}

% \section*{Supplementary Information}
\cleardoublepage
\setcounter{figure}{0}
\setcounter{page}{1}
\renewcommand\thefigure{S.\arabic{figure}}% 

\label{Supp}
% \break
%%%%%%%%%%%%%%%%%%%%%%%%%%%%%%%%%%%%
%%%%%%%%%%%%%%%%%%%%%%%%%%%%%%%%%%%%
%%%%%%%%%%%%%%%%%%%%%%%%%%%%%%%%%%%%
%%%%%%%%%%%%%%%%%%%%%%%%%%%%%%%%%%%%
%%%%%%%%%%%%%%%%%%%%%%%%%%%%%%%%%%%%
\begin{figure*}[bt!]
\centering
	\includegraphics[width=0.9\textwidth]{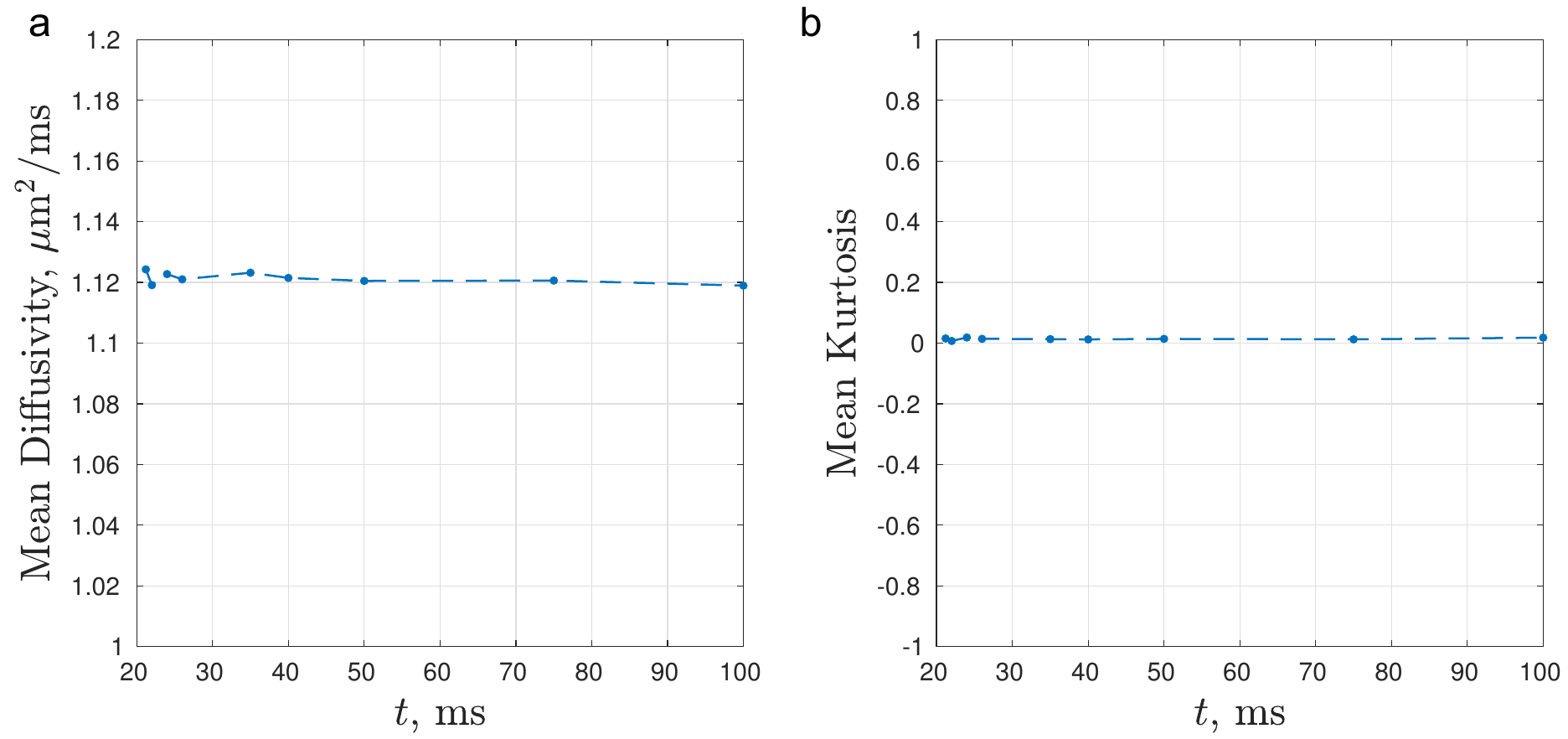}
	\caption{Ice-water phantom diffusivity and kurtosis fitted values with respect to diffusion time at temperature $=0^{\circ}$C. No time-dependence was observed as expected, eliminating possible pulse sequence contributions to the time-dependence observed in cortical gray matter.  }
	\label{Fig:Ice_waterphantom}
\end{figure*}
\renewcommand{\headrulewidth}{0.5pt}
\fancyhead[LH]{Supplementary Material}
\fancyhead[RH]{Lee and Papaioannou, et al., NeuroImage, 2020}
%%%%%%%%%%%%%%%%%%%%%%%%%%%%%%%%%%%%
%%%%%%%%%%%%%%%%%%%%%%%%%%%%%%%%%%%%
%%%%%%%%%%%%%%%%%%%%%%%%%%%%%%%%%%%%
%%%%%%%%%%%%%%%%%%%%%%%%%%%%%%%%%%%%
%%%%%%%%%%%%%%%%%%%%%%%%%%%%%%%%%%%%
\begin{figure*}[bt!]
\centering
	\includegraphics[width=0.9\textwidth]{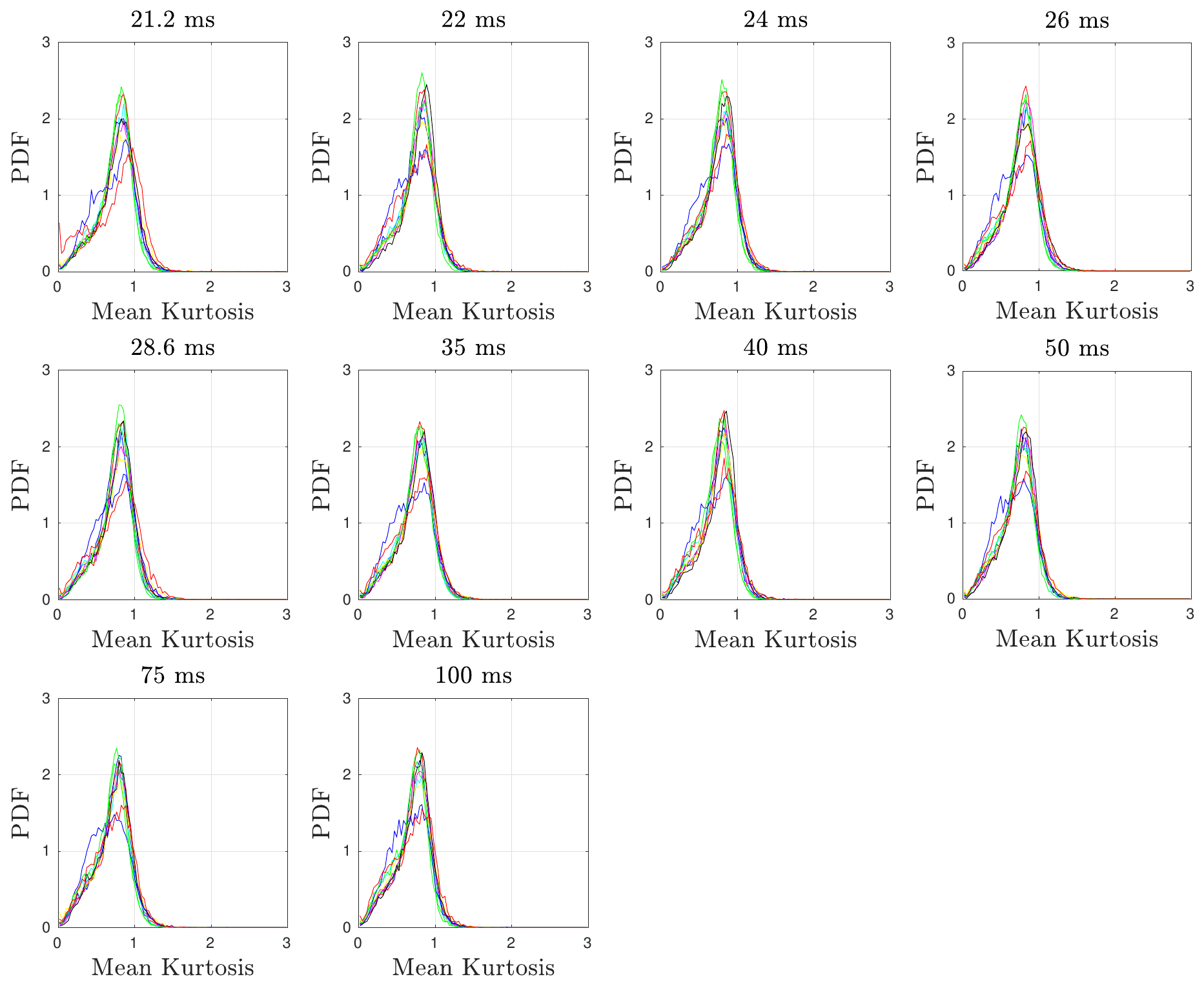}
	\caption{Histograms of the mean kurtosis for each subject and time-point studied in this work. The histogram suggests reasonable inter-subject variability.}
	\label{fig:suppl-MK}
\end{figure*}
%%%%%%%%%%%%%%%%%%%%%%%%%%%%%%%%%%%%%
%%%%%%%%%%%%%%%%%%%%%%%%%%%%%%%%%%%%%
%%%%%%%%%%%%%%%%%%%%%%%%%%%%%%%%%%%%%
%%%%%%%%%%%%%%%%%%%%%%%%%%%%%%%%%%%%%
%%%%%%%%%%%%%%%%%%%%%%%%%%%%%%%%%%%%%

\begin{figure*}[bt!]
\centering
	\includegraphics[width=0.35\textwidth]{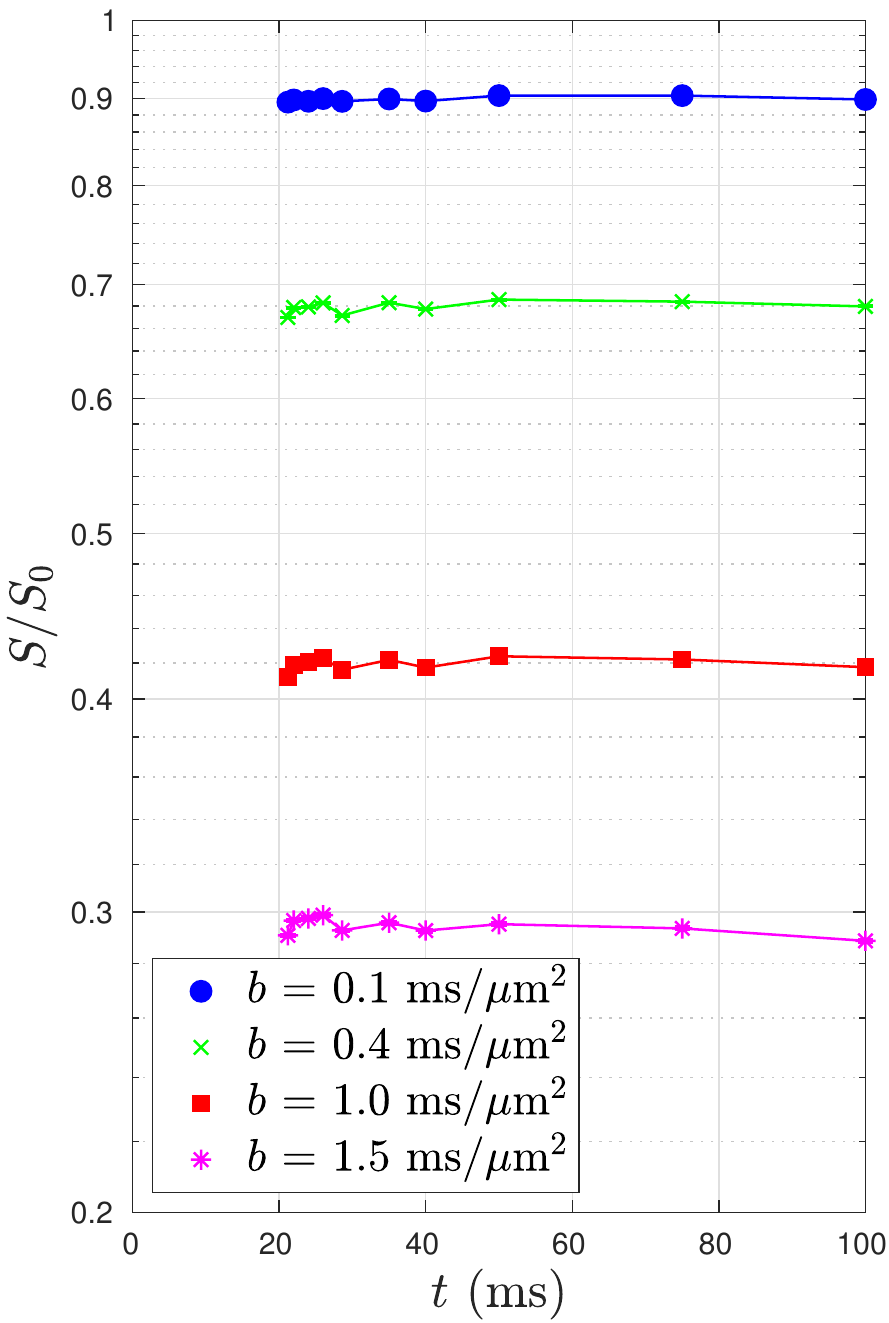}
	\caption[]{\mnotes{R2.6}Normalized diffusion signals in cortical gray matter averaged over 10 healthy subjects, with respect to diffusion weighting $b$ and diffusion time $t$. At low $b$, diffusion signals at different times are similar; on the other hand, at high $b$, diffusion signals have relatively larger changes over diffusion times. Therefore, the observed diffusivity time-dependence (mainly estimated from low $b$ data) is relatively weak and noisy, whereas the observed kurtosis time-dependence (mainly estimated from high $b$ data) is significant.}
	\label{fig:suppl-S-b}
\end{figure*}

\begin{figure*}[bt!]
\centering
	\includegraphics[width=0.99\textwidth]{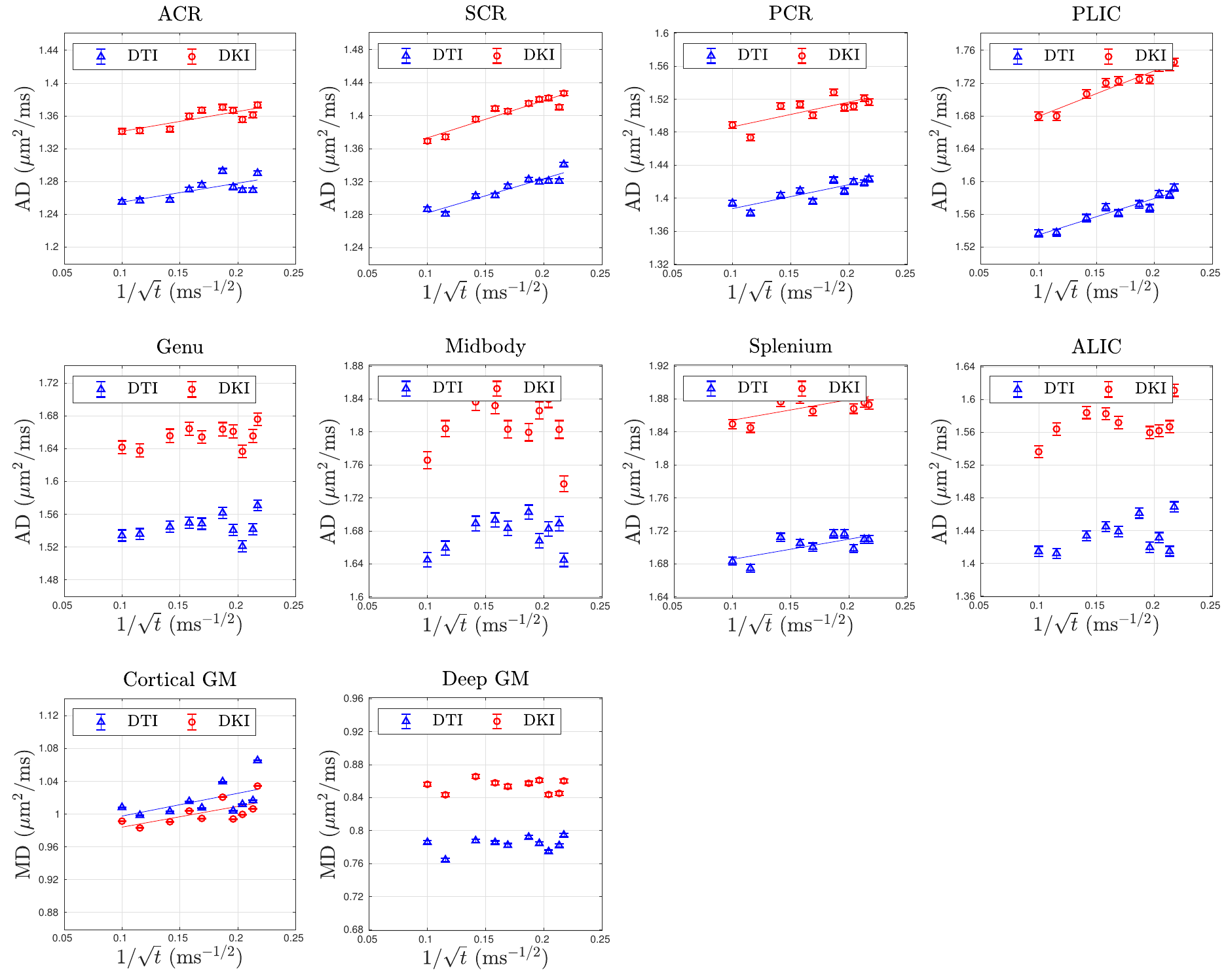}
	\caption[]{\mnotes{R2.5}Axial diffusivity (AD) time-dependence, estimated using either diffusion tensor imaging (DTI) (maximal b-value $\leq$ 0.4 ms/$\mu$m$^2$) or DKI, in white matter (WM) ROIs averaged over 10 healthy subjects. Solid lines are fits for significant time-dependence (P-value $<$ 0.05), $\text{AD}(t)\simeq D_\infty+c_D/\sqrt{t}$  \citep{novikov2014revealing,Fieremans2016}. (ACR/SCR/PCR = anterior/superior/posterior corona radiata, PLIC/ALIC = posterior/anterior limb of the internal capsule, Genu/Midbody/Splenium of corpus callosum). As a reference, mean diffusivity (MD) time-dependence, estimated using either DTI or DKI, in cortical and deep gray matter (GM) ROIs averaged over 10 healthy subjects is also shown. Solid lines are fits to \eqref{c'-theta} for significant time-dependence (P-values $<$ 0.05).}
	\label{fig:suppl-AD-MD}
\end{figure*}

\begin{figure*}[bt!]
\centering
	\includegraphics[width=0.99\textwidth]{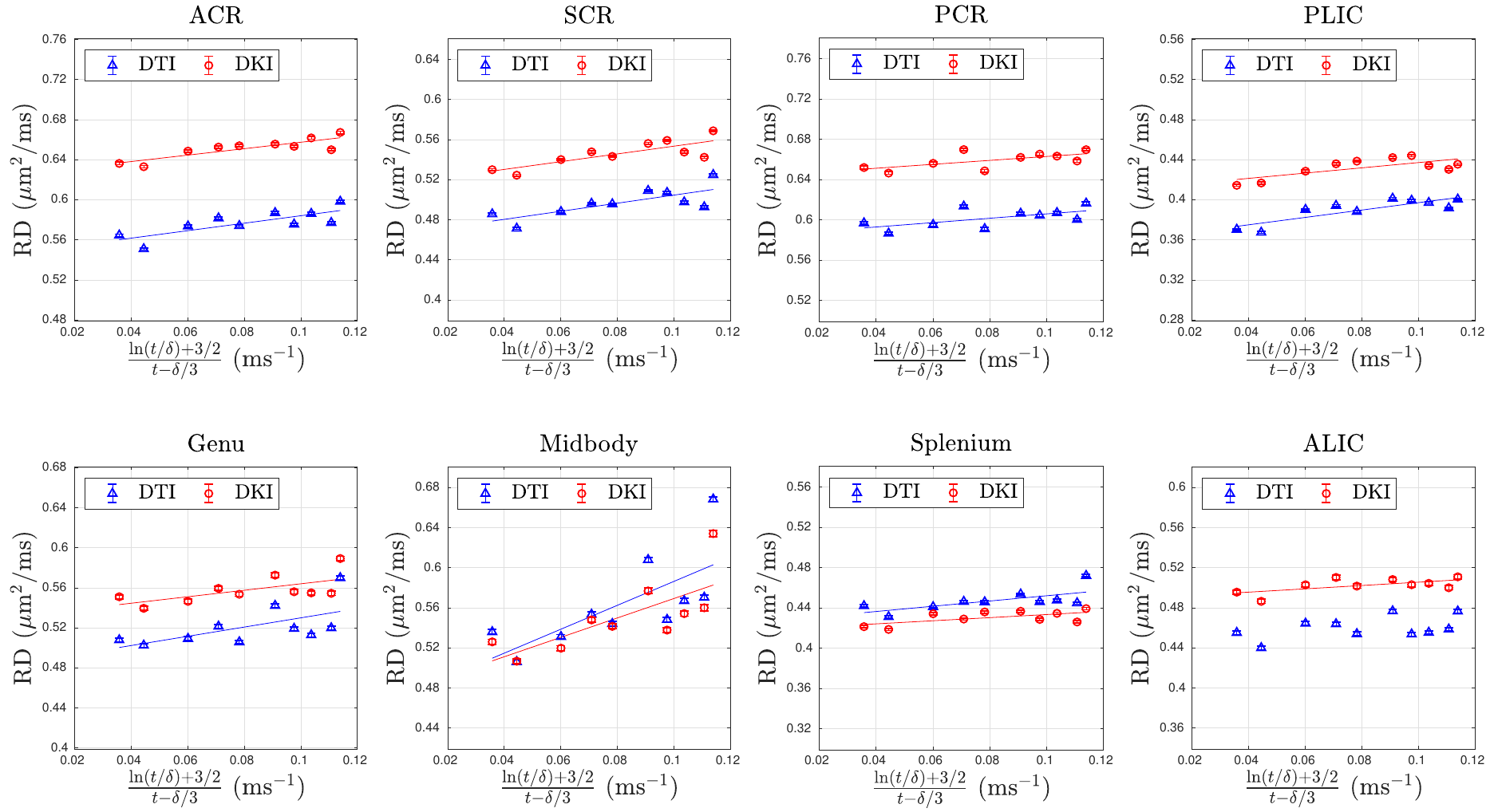}
	\caption[]{\mnotes{R2.5}Radial diffusivity (RD) time-dependence, estimated using either diffusion tensor imaging (DTI) (maximal b-value $\leq$ 0.4 ms/$\mu$m$^2$) or DKI, in white matter (WM) ROIs averaged over 10 healthy subjects. Solid lines are fits for significant time-dependence (P-value $<$ 0.05), $\text{RD}(t)\simeq D_\infty + c_D\cdot\left[\ln(t/\delta)+3/2\right]/(t-\delta/3)$ \citep{Burcaw2015,Fieremans2016}. (ACR/SCR/PCR = anterior/superior/posterior corona radiata, PLIC/ALIC = posterior/anterior limb of the internal capsule, Genu/Midbody/Splenium of corpus callosum)}
	\label{fig:suppl-RD}
\end{figure*}

\end{document}